\documentclass{article}

\usepackage{PRIMEarxiv}

\usepackage[utf8]{inputenc} % allow utf-8 input
\usepackage[T1]{fontenc}    % use 8-bit T1 fonts
\usepackage{hyperref}       % hyperlinks
\usepackage{url}            % simple URL typesetting
\usepackage{booktabs}       % professional-quality tables
\usepackage{amsfonts}       % blackboard math symbols
\usepackage{nicefrac}       % compact symbols for 1/2, etc.
\usepackage{microtype}      % microtypography
\usepackage{lipsum}
\usepackage{fancyhdr}       % header
\usepackage{graphicx}       % graphics
\graphicspath{{media/}}     % organize your images and other figures under media/ folder

\usepackage[round]{natbib}
\usepackage{amsfonts}
\usepackage{amsmath}
\usepackage{algorithm}
\usepackage{algpseudocode}

\newcommand{\btheta}{ {\boldsymbol{\theta}} }
\newcommand{\bphi}{ {\boldsymbol{\phi}} }

\usepackage{enumitem}

\title{Latent Diffusion Model for Conditional Reservoir Facies Generation}

\author{
  Daesoo Lee \\
  Norwegian University of Science and Technology \\
  \And
  Oscar Ovanger \\
  Norwegian University of Science and Technology \\
  \And
  Jo Eidsvik \\
  Norwegian University of Science and Technology \\
  \And
  Erlend Aune \\ 
  Norwegian University of Science and Technology \\
  BI Norwegian Business School \\
  Abelee \\
  \And
  Jacob Skauvold \\
  Norwegian Computing Center \\
  \And
  Ragnar Hauge \\
  Norwegian Computing Center
}

\begin{document}
\maketitle
% \def\thefootnote{*}\footnotetext{These authors contributed equally to this work.}
% \def\thefootnote{\arabic{footnote}}
% text text text\footnote{normal footnote}

\begin{abstract}
Creating accurate and geologically realistic reservoir facies based on limited measurements is crucial for field development and reservoir management, especially in the oil and gas sector. Traditional two-point geostatistics, while foundational, often struggle to capture complex geological patterns. Multi-point statistics offers more flexibility, but comes with its own challenges related to pattern configurations and storage limits. With the rise of Generative Adversarial Networks (GANs) and their success in various fields, there has been a shift towards using them for facies generation. However, recent advances in the computer vision domain have shown the superiority of diffusion models over GANs. Motivated by this, a novel Latent Diffusion Model is proposed, which is specifically designed for conditional generation of reservoir facies. The proposed model produces high-fidelity facies realizations that rigorously preserve conditioning data. It significantly outperforms a GAN-based alternative. 
Our implementation on GitHub: \url{https://github.com/ML4ITS/Latent-Diffusion-Model-for-Conditional-Reservoir-Facies-Generation}.
\end{abstract}

% keywords can be removed
% \keywords{First keyword \and Second keyword \and More}
% \keywords{Conditional facies generation \and Reservoir facies \and Diffusion model \and Latent Diffusion Model}

\section{Introduction}
\label{intro}

Creating accurate and geologically realistic reservoir facies predictions based on limited measurements is a critical task in development and production of oil and gas resources. It is also very relevant in connection with CO$_2$ storage, where one makes decisions about injection strategies to manage leakage risk and ensure safe long-term operations. In both contexts, key operational decisions are based on realizations of stochastic reservoir models. Through the use of multiple realizations, one can go beyond point-wise prediction of facies, and additionally quantify spatial variability and correlation. This gives better descriptions of the relevant heterogeneity.

When generating facies realizations, one must honor both geological knowledge and reservoir-specific data. A wide range of stochastic models have been proposed to solve this problem. A good overview can be found in the book by \cite{pyrcz2014geostatistical}. There are variogram-based models, where the classical concept of a variogram-based Gaussian field \citep[see for instance][]{cressie2015statistics} is combined with a discretization scheme to generate facies. Then there are more geometric approaches, such as object models or process-mimicking models, where facies are described as geometric objects with an expected shape and uncertainty. Of particular interest here are multiple-point models, which use a training image to generate a pattern distribution, and then generate samples following this distribution.

Multiple-point models are very flexible, and allow for complex interactions between any number of facies. But as the method fundamentally hinges on storing pattern counts, there are strict limitations due to memory. In practice, only a limited number of patterns can be handled, leading to restrictions in pattern size and a demand for stationarity of patterns. Furthermore, the simulation algorithm has clear limitations in its ability to reproduce the patterns, so a realization will typically contain many patterns not found in the initial database, leading to unwanted geometries \citep{zhang2019generating}. Limitations like these have led to the adoption of models such as generative adversarial networks \citep[GANs,][]{goodfellow2020generative}. In recent years, GANs have gained substantial attention for the conditional generation of realistic facies while retaining conditional data in a generated sample, see e.g.   \cite{chan2019parametric,zhang2019generating,azevedo2020generative,pan2021stochastic,song2021gansim,zhang2021u,yang2022automatic,razak2022conditioning,hu2023multi}. 

\begin{figure} %
  \centering
  \includegraphics[width=0.67\textwidth]{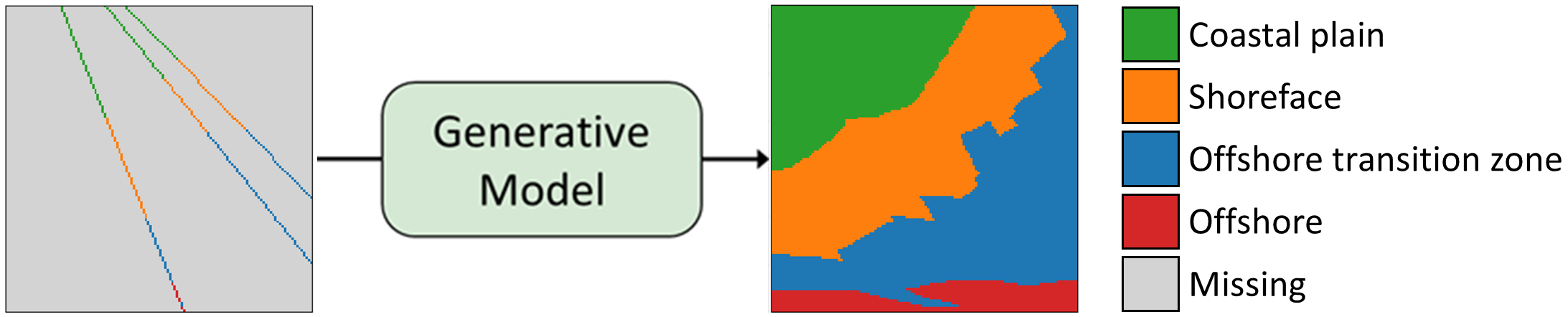}
  \caption{Illustration of our conditional reservoir generation problem in which the generative model stochastically samples a realistic reservoir (right) given the limited measurements (left).
  % In our dataset, three different facies are present, colored in beige, blue, and brown, respectively, while the non-facies regions are marked in green. 
  The regions with no information are denoted in grey. 
  }
  \label{fig:problem_illustration}
\end{figure}

We frame stochastic facies modeling as a conditional generation problem in machine learning. This view is motivated by the observation that in some existing methods for reservoir modeling, generating unconditional realizations is comparatively easy, and the difficulty increases sharply as one moves to generating conditional realizations. The principal idea of this paper is to exploit this difficulty gap by using easily generated unconditional realizations as training data for a machine learning model. Crucially, this model will learn not only how to reproduce features seen in the training realizations, but also how to honor conditioning data. A model successfully trained in this way can generate conditional realizations given previously unseen conditioning data.
Fig.~\ref{fig:problem_illustration} illustrates our conditional generation problem.

Recent studies in computer vision have demonstrated the superiority of diffusion models over GANs in terms of generative performance \citep{dhariwal2021diffusion,rombach2022high,ho2022cascaded,kim2022diffusionclip}. As a result, diffusion models are state-of-the-art for image generation, while the popularity of GANs has diminished due to limitations including convergence problems, mode collapse, generator-discriminator imbalance, and sensitivity to hyperparameter selection. Latent diffusion models (LDMs) are a type of diffusion model in which the diffusion process occurs in a latent space rather than in pixel space \citep{rombach2022high}. LDMs have become popular because they combine computational efficiency with good generative performance.

Motivated by the progress made with diffusion models on computer vision and image processing tasks, this work proposes a novel LDM, specifically designed for the generation of conditional facies realizations in a reservoir modeling context. Its appeal lies in the ability to strictly preserve conditioning data in the generated realizations. To the authors' knowledge, this is the first work to adopt a diffusion model for conditional facies generation.

Experiments were carried out using a dataset of 5,000 synthetic 2D facies realizations to evaluate the proposed diffusion model against a GAN-based alternative. The diffusion model achieved robust conditional facies generation performance in terms of fidelity, sample diversity, and the preservation of conditional data, while the GAN-based model struggled with multiple critical weaknesses.

To summarize, the contributions of this paper are:
\begin{itemize}
    \vspace{-0.2cm}\item the adoption of a diffusion model for conditional facies generation,
    \vspace{-0.2cm}\item a novel LDM, designed to preserve observed facies data in generated samples,
    \vspace{-0.2cm}\item conditional facies generation with high fidelity, sample diversity, and robust preservation.
\end{itemize}

In Section \ref{sec:Relawork}, we describe GANs and background information for the LDMs. In Section \ref{sec:method}, we present our suggested methodology for conditional facies realizations with LDMs. In Section \ref{sec:results}, we show experimental results of our method applied to a bedset model with stacked facies, including the comparison with GANs. In Section \ref{sec:concl}, we summarize and discuss future work.

\section{Background on Generative Models}
\label{sec:Relawork}
\subsection{Generative Adversarial Network for Conditional Image Generation} 
GANs were a breakthrough innovation in the field of generative AI when they emerged in 2014. The core mechanism of GANs involves two neural networks, a generator and a discriminator, engaged in a sort of cat-and-mouse game. The generator aims to mimic the real data, while the discriminator tries to distinguish between real and generated data. Through iterative training, the generator improves its ability to create realistic data, and the discriminator becomes more adept at identifying fakes.

\begin{figure}%
  \centering
  \includegraphics[width=0.45\textwidth]{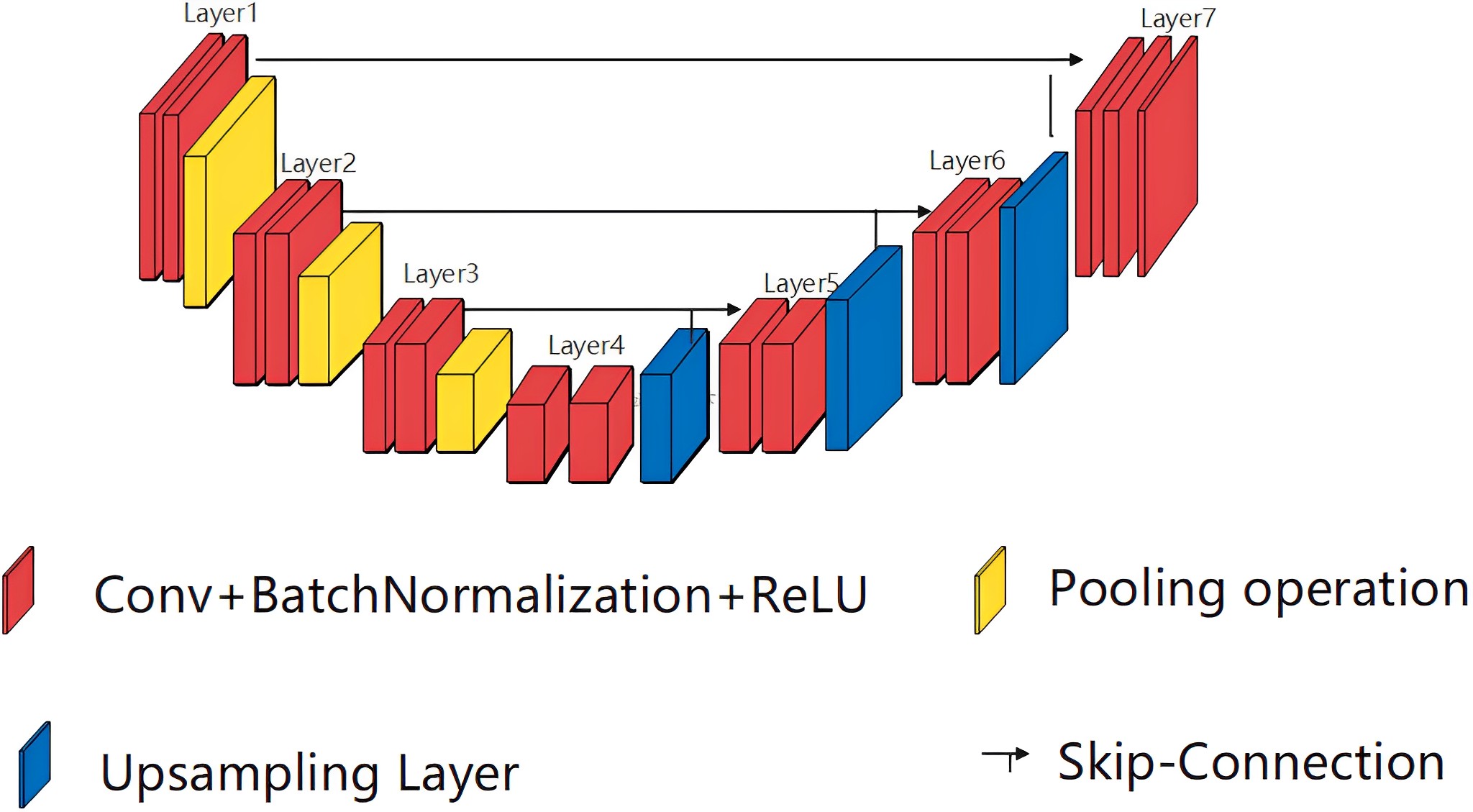}
  \caption{Illustration of the U-Net architecture \citep{cai2022novel}, where Conv denotes a convolutional layer.
  U-Net is a convolutional neural network architecture, featuring an encoder (first half of U-Net) and decoder (second half) structure with skip connections that allow for the transfer of spatial information across layers, which in turn enables precise localization and high-resolution output. 
  }
  \label{fig:unet_illustration}
\end{figure}

Conditional Generative Adversarial Networks (CGANs) were proposed by \cite{mirza2014conditional} in the same year as the GAN. The CGAN was designed to guide the image generation process of the generator given conditional data such as class labels and texts as auxiliary information.
Since then, CGANs have been further developed to perform various tasks. 
Among these, \cite{isola2017image} stands out from the perspective of conditional facies generation, proposing a type of CGAN called Pixel2Pixel (Pix2Pix), which has become a popular GAN method for image-to-image translation. Pix2Pix works by training a CGAN to learn a mapping between input images and output images from different distributions. For instance, the input could be a line drawing, and the output a corresponding color image. The mapping can be realized effectively with the help of the U-Net architecture \citep{ronneberger2015u}, illustrated in Fig.~\ref{fig:unet_illustration}.

Image-to-image translation is directly relevant to conditional facies generation because the input can be facies observations on a limited subset of the model domain, and the output can be a complete facies model. This is the typical situation when the goal is to generate 2D or 3D facies realizations from sparse facies observations at the well locations.

\subsection{GANs for Conditional Facies Generation}
\cite{dupont2018generating} were the first to adopt a GAN for conditional facies generation, overcoming the limitations of traditional geostatistical methods by producing varied and realistic geological patterns that honor measurements at data points. However, the latent vector search required to ensure a match with the conditioning data makes the sampling process inefficient. \cite{chan2019parametric} introduced a second inference network that enables the direct generation of realizations conditioned on observations, thus providing a more efficient conditional sampling approach. \cite{zhang2019generating} introduced a GAN-based approach to generate 3D facies realizations, specifically focusing on complex fluvial and carbonate reservoirs. Their paper clearly demonstrated the superiority of GAN over MPS for this application. \cite{azevedo2020generative} used GANs in a similar way, but the evaluation of its conditional generation makes this study different from others. Where GANs from other studies typically condition on multiple sparse points, the paper considered conditioning on patches and lines. Because such shapes typically involve a larger region than multiple sparse points, their conditional setup is more difficult, which is demonstrated in experiments. \cite{pan2021stochastic} used Pix2Pix, adopting the U-Net architecture. It takes a facies observation and noise as input, and then stochastically outputs a full facies realization. Notably, the preservation of conditional data in a generated sample was shown to be effective due to the U-Net architecture that enables precise localization. A paper by \cite{zhang2021u} is concurrent with and methodologically similar to \cite{pan2021stochastic} as both articles propose a GAN built on U-Net. However, the U-Net GAN of \cite{zhang2021u} has an additional loss term to ensure sample diversity, which simplifies the sampling process. Subsequently, many studies have sought to improve conditional facies generation using GANs, working within the same or similar frameworks as the studies mentioned above \cite{song2021gansim,yang2022automatic,razak2022conditioning,hu2023multi}.

The main difference between the current study and previous research is the type of generative model employed, specifically the choice of a diffusion model over a GAN. This also leads to a specific network architecture used to enable conditioning. Another difference is that whereas much earlier work is done in a top-down view, we focus on a vertical section. A consequence of this is that we get a different structure for the conditioning data. In a vertical section, well data become paths, giving connected lines of cells with known facies. In the top-down view, wells appear as scattered individual grid cells with known facies.

\subsection{Denoising Diffusion Probabilistic Model (DDPM)} 

\begin{figure}%
  \centering
  \includegraphics[width=0.7\textwidth]{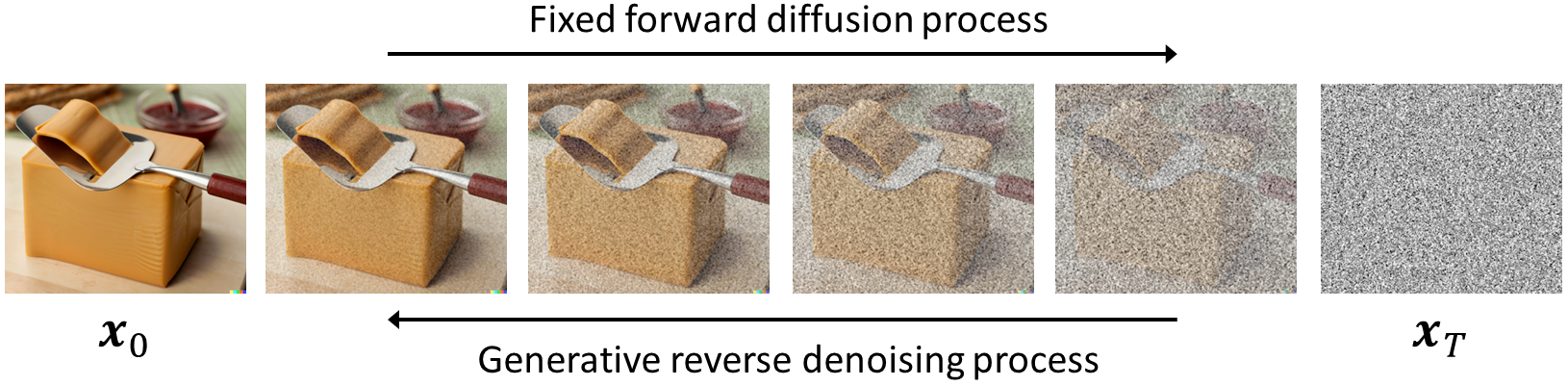}
  \caption{Illustration of the principle of a diffusion process. The diffusion modeling mainly consists of 1) forward process (noising) and 2) reverse process (denoising).
  The noising process begins with a data sample and incrementally adds Gaussian noise over multiple time steps to convert it into a Gaussian noise sample; conversely, the denoising process iteratively refines this Gaussian noise sample back into a data-like sample, guided by a neural network trained specifically for this denoising task.}
  \label{fig:ddpm}
\end{figure}

\cite{ho2020denoising} represented a milestone for diffusion model-based generative modeling. %Their suggested DDPM is a prominent approach in the field of generative image modeling that has gained much attention in recent years. 
DDPMs offer a powerful framework for generating high-quality image samples from complex data distributions. 
At its core, a DDPM leverages the principles of diffusion processes to model a data distribution. It operates by iteratively denoising a noisy sample and gradually refining it to generate a realistic sample as illustrated in Fig.~\ref{fig:ddpm}. This denoising process corresponds to the reverse process of a fixed Markov process of a certain length.

% {\bf{Jo's comment, here the forward diffusion process goes right-to-left, below it goes left-to-right. Not that it matters, but consistency could make the paper / figures easier to read.}}

A DDPM employs a denoising autoencoder, denoted by $\epsilon_\theta(\textit{\textbf{x}}_t, t); t=1, \dots, T$. The denoising autoencoder gradually refines the initial noise $\textit{\textbf{x}}_T$ to generate a high-quality sample $\textit{\textbf{x}}_0$ that closely resembles the target data distribution. 
A U-Net is used for the denoising autoencoder since its architecture provides effective feature extraction, preservation of spatial details, and robust performance in modeling complex data distributions \citep{baranchuk2021label}. 

\paragraph{Training: Prediction of Noise in $\textit{\textbf{x}}_t$}
DDPM training consists of two key components: the non-parametric forward process and the parameterized reverse process. The former component represents the gradual addition of Gaussian noise. In contrast, the reverse process needs to be learned to predict noise $\boldsymbol{\epsilon}$ in $\textit{\textbf{x}}_t$. 
% This learning process involves training a sequence of weight-shared denoising autoencoders within DDPM. These denoising autoencoders are responsible for predicting noise $\epsilon$ in $x_t$. 
Its loss function is defined by 
\begin{equation} 
\label{eq:L_DDPM}
L_{\mathrm{DDPM}} = \mathbb{E}_{\textit{\textbf{x}}, \epsilon \sim \mathcal{N}(\textbf{0},\textbf{I}),t} \left[ \| \boldsymbol{\epsilon} - \epsilon_{\btheta}(\textit{\textbf{x}}_t, t)  \|_2^2 \right],
\end{equation}
where $\epsilon_{\btheta}$ denotes the denoising autoencoder with parameters $\btheta$. Eq.~\eqref{eq:L_DDPM} measures the discrepancy between the noise and the predicted noise by the denoising autoencoder. 
% The loss function serves as a guidance signal for the model to optimize its denoising capability. By minimizing this loss, DDPM learns to effectively denoise $x_t$, progressively reducing the noise level during the reverse process. 
While Eq.~\eqref{eq:L_DDPM} defines a loss function for unconditional generation, the loss for conditional generation is specified by
\begin{equation} 
\label{eq:L_DDPM_c}
L_{\mathrm{DDPM},c} = \mathbb{E}_{\textit{\textbf{x}}, c, \boldsymbol{\epsilon} \sim \mathcal{N}(\textbf{0},\textbf{I}),t} \left[ \| \boldsymbol{\epsilon} - \epsilon_{\btheta}(\textit{\textbf{x}}_t, t, c)  \|_2^2 \right],
\end{equation}
where $c$ denotes conditional data such as texts or image class, and in our situation, the observed facies classes in wells. Typically, $L_{\mathrm{DDPM}}$ and $L_{\mathrm{DDPM},c}$ are both minimized during training,  to allow both unconditional and conditional generation. 
For details, see \cite{ho2022classifier}.

\paragraph{Sampling via Learned Reverse Process}
The forward diffusion process is defined as $q(\textit{\textbf{x}}_t | \textit{\textbf{x}}_{t-1}) = \mathcal{N}(\textit{\textbf{x}}_t; \sqrt{1 - \beta_t} \textit{\textbf{x}}_{t-1}, \beta_t \textbf{I})$ where $\beta_t$ is called a variance schedule and $1 \geq \beta_T > \beta_1 \geq 0$. Equivalently, it can be written $\textit{\textbf{x}}_t = \sqrt{1 - \beta_t} \textit{\textbf{x}}_{t-1} + \sqrt{\beta_t} \boldsymbol{\epsilon}_{t-1}$ with $\boldsymbol{\epsilon}_{t-1} \sim \mathcal{N}(\textbf{0}, \textbf{I})$. We further reformulate the equation with respect to $\textit{\textbf{x}}_{t-1}$ and it becomes
\begin{align}
\label{eq:x_{t-1}}
\textit{\textbf{x}}_{t-1} = (\textit{\textbf{x}}_t - \sqrt{\beta_t} \boldsymbol{\epsilon}_{t-1}) / \sqrt{1 - \beta_t} = (\textit{\textbf{x}}_t - \sqrt{1 - \alpha_t} \boldsymbol{\epsilon}_{t-1}) / \sqrt{\alpha_t},
\end{align}
where $\alpha_t = 1 - \beta_t$. Then we can go backwards, sampling $\textit{\textbf{x}}_0$ from $\textit{\textbf{x}}_T$ by recursively applying Eq.~\eqref{eq:x_{t-1}} for $t=T, \dots, 2, 1$.

% explain latent diffusion model (LDM)
    % p(z|null), p(z|c)
    % show LDM in a simple figure including its conditioning.
\subsection{Latent Diffusion Model (LDM)}
\label{sect:Latent Diffusion Model}
\begin{figure}%
  \centering
  \includegraphics[width=0.7\textwidth]{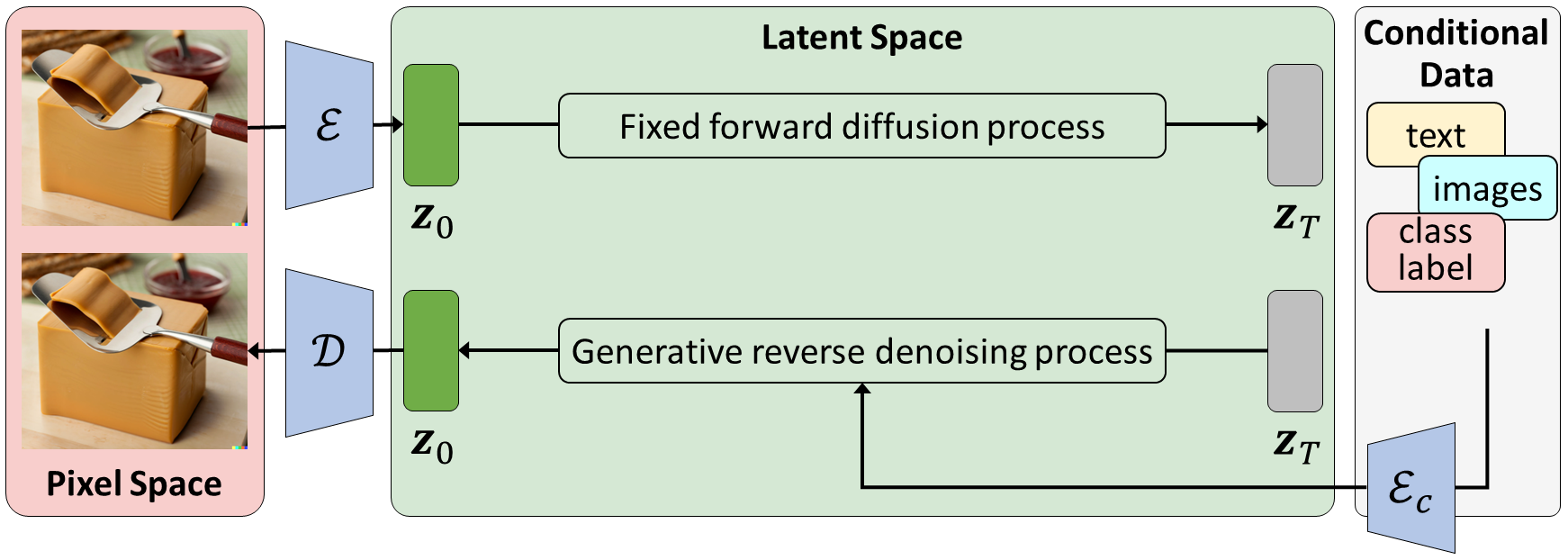}
  \caption{Overview of LDM. 
  The encoder $\mathcal{E}$ and decoder $\mathcal{D}$ enable data compression, enabling the forward and reverse processes to operate in a reduced-dimensional space. This eases the task of learning prior and posterior distributions and improves computational efficiency. 
  In addition, conditional data can be fed into the reverse process, enabling conditional generation.
  }
  \label{fig:ldm}
\end{figure}

\begin{figure}%
  \centering
  \includegraphics[width=0.7\textwidth]{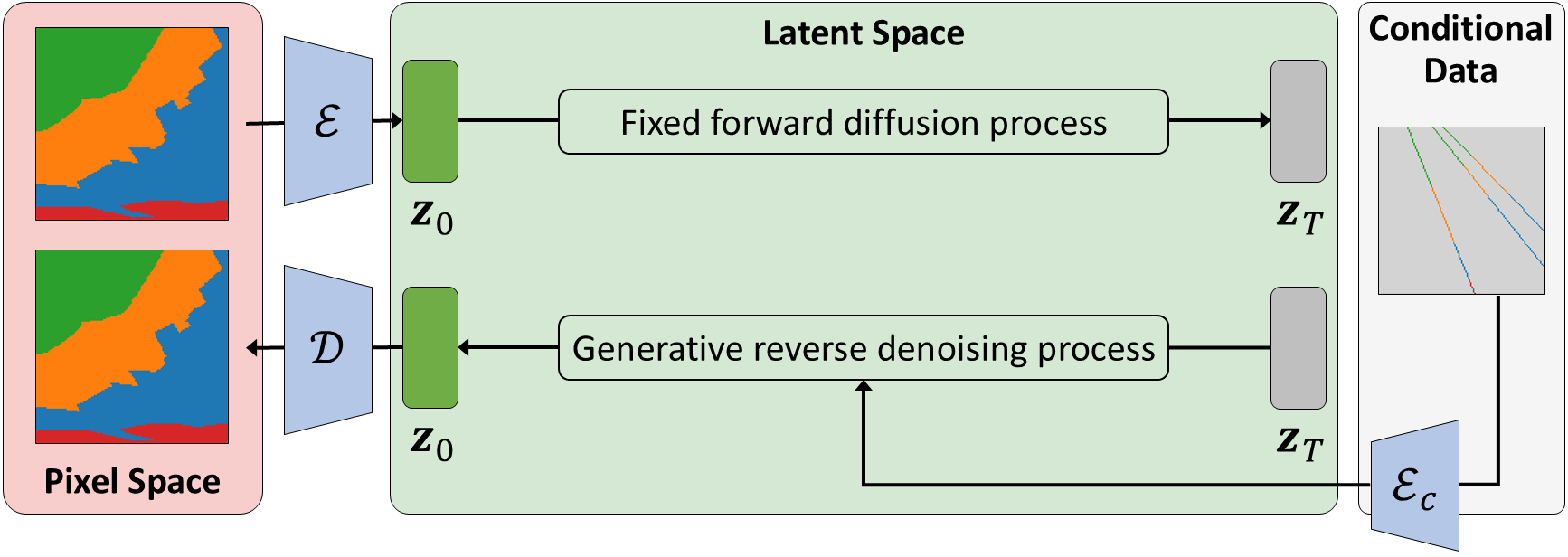}
  \caption{Overview of our proposed method. Our method can be regarded as an adapted version of LDM to effectively handle the categorical input and allow maximal preservation of conditional facies data in generated facies while maintaining the high fidelity of generated facies.
  }
  \label{fig:ldm-to-our_problem}
\end{figure}

LDMs extend DDPMs by introducing a diffusion process in a latent space. The main idea of LDMs is illustrated in Fig.~\ref{fig:ldm}, aligning with the overview of our proposed method for conditional facies generation as depicted in Fig.~\ref{fig:ldm-to-our_problem}. In the common situation, data are typically text or images as indicated to the far right in Fig.~\ref{fig:ldm}. In our setting, the conditional data are facies observations along a few well paths in the subsurface. 

Compared with a DDPM, an LDM has two additional components: encoder $\mathcal{E}$ and decoder $\mathcal{D}$. 
The encoder transforms $\textit{\textbf{x}}$ into a latent representation, $\textit{\textbf{z}} = \textit{\textbf{z}}_0 = \mathcal{E}(\textit{\textbf{x}})$, while the decoder reconstructs $\textit{\textbf{z}} $ to produce $\tilde{\textit{\textbf{x}}}$.
Importantly, the encoding and decoding processes involve downsampling and upsampling operations, respectively. The encoder and decoder are trained so that $\tilde{\textit{\textbf{x}}}$ is as close as possible to $\textit{\textbf{x}}$. This is ensured by minimizing a reconstruction loss between $\textit{\textbf{x}}$ and $\tilde{\textit{\textbf{x}}}$. 
Notably, the forward and backward processes are now taking place in the latent space, therefore $\textit{\textbf{z}}_T$ denotes a Gaussian noise sample. In Fig.~\ref{fig:ldm}, $\mathcal{E}_c$ denotes an encoder for conditional data. The encoded conditional data is fed into the reverse process for conditioning the generation process.%, as formulated in Eq.~\eqref{eq:L_LDM_c-2}. 

% \begin{figure}
%   \centering
%   \includegraphics[width=0.7\textwidth]{.png}
%   \caption{Example of the denoising process using our proposed method .
%   }
%   \label{fig:ldm_denoising_example}
% \end{figure}

The main advantage of LDMs over DDPMs is computational efficiency. The encoder $\mathcal{E}$ compresses high-dimensional data $\textit{\textbf{x}}$ into a lower-dimensional latent space represented via latent variable $\textit{\textbf{z}}$. This dimensionality reduction significantly reduces the computational cost, making LDM more feasible to be trained on local devices. However, a trade-off exists between computational efficiency and sample quality. Increasing the downsampling rate of $\mathcal{E}$ increases the computational efficiency but typically results in a loss of sample quality, and vice versa.

Training of LDMs adopts a two-staged modeling approach \citep{van2017neural,chang2022maskgit}. The first stage (stage~1) is for learning the compression and decompression of $\textit{\textbf{x}}$ by training $\mathcal{E}$ and $\mathcal{D}$, and the second stage (stage~2) is for learning the prior and posterior distributions by training $\epsilon_{\btheta}$.

In stage~1, $\textit{\textbf{x}}$ is encoded into $\textit{\textbf{z}}$ and decoded back into the data space. The training of $\mathcal{E}$ and $\mathcal{D}$ is conducted by minimizing the following reconstruction loss:
% The training of $\mathcal{E}$ and $\mathcal{D}$ involves minimizing the reconstruction loss. In LDM, the $L_2$ reconstruction loss $\| x -  \tilde{x}\|_2^2$ is used,
\begin{align} 
\label{eq:LDM_stage1_loss}
\| \textit{\textbf{x}} -  \mathcal{D}(\mathcal{E}(\textit{\textbf{x}}))\|_2^2.
\end{align}

In stage~2, the denoising autoencoder $\epsilon_{\btheta}$ is trained to learn prior and posterior distributions, while $\mathcal{E}$ and $\mathcal{D}$ are set to be untrainable (frozen). This involves minimizing
\begin{align} 
\label{eq:L_LDM}
L_{\mathrm{LDM}} &= \mathbb{E}_{\mathcal{E}(\textit{\textbf{x}}), \boldsymbol{\epsilon} \sim \mathcal{N}(\textbf{0},\textbf{I}),t} \left[ \| \boldsymbol{\epsilon} - {\epsilon}_{\btheta}(\textit{\textbf{z}}_t, t)  \|_2^2 \right], \hspace{19mm} \mbox{Prior training}, \\
\label{eq:L_LDM_c}
L_{\mathrm{LDM},c} &= \mathbb{E}_{\mathcal{E}(\textit{\textbf{x}}), c, \boldsymbol{\epsilon} \sim \mathcal{N}(\textbf{0},\textbf{I}),t} \left[ \| \boldsymbol{\epsilon} - {\epsilon}_{\btheta}(\textit{\textbf{z}}_t, t, c)  \|_2^2 \right], \hspace{13mm} \mbox{Posterior training}.
\end{align}

% In this paper, we propose our method tailored for conditional reservoir generation based on LDM for its computational efficiency and feasibility.

The recent work that proposed DALLE-2 \citep{ramesh2022hierarchical} empirically found that predicting $\textit{\textbf{z}}_0$ instead of $\boldsymbol{\epsilon}$ results in better training. We adopt the same approach for better training and methodological simplicity of our conditional sampling. 
Hence, we in stage~2 instead minimize
\begin{align} 
\label{eq:L_LDM-2}
L_{\mathrm{LDM}}(\textit{\textbf{z}}, g_{\btheta}) &= \mathbb{E}_{\mathcal{E}(\textit{\textbf{x}}), \boldsymbol{\epsilon} \sim \mathcal{N}(\textbf{0},\textbf{I}),t} \left[ \| \textit{\textbf{z}}_0 - g_{\btheta}(\textit{\textbf{z}}_t, t)  \|_2^2 \right], \\
\label{eq:L_LDM_c-2}
L_{\mathrm{LDM},c}(\textit{\textbf{z}}, c, g_{\btheta}) &= \mathbb{E}_{\mathcal{E}(\textit{\textbf{x}}), c, \boldsymbol{\epsilon} \sim \mathcal{N}(\textbf{0},\textbf{I}),t} \left[ \| \textit{\textbf{z}}_0 - g_{\btheta}(\textit{\textbf{z}}_t, t, c)  \|_2^2 \right],
\end{align}
where $g_{\btheta}$ is a denoising autoencoder that predicts $\textit{\textbf{z}}_0$ instead of $\boldsymbol{\epsilon}$. 
Then sampling in the latent space can be formulated as
%utilizing $q(\textit{\textbf{x}}_{t-1} | \textit{\textbf{x}}_t, \textit{\textbf{x}}_0)$, as formulated in Equation~(6) of \cite{ho2020denoising}. We can reformulate it with respect to $\textit{\textbf{z}}$ and it becomes 
$q(\textit{\textbf{z}}_{t-1} | \textit{\textbf{z}}_t, \textit{\textbf{z}}_0) = \mathcal{N}(\textit{\textbf{z}}_{t-1}; \tilde{\mu}_t(\textit{\textbf{z}}_t, \textit{\textbf{z}}_0), \tilde{\beta}_t \textbf{I})$ where $\tilde{\mu}_t(\textit{\textbf{z}}_t, \textit{\textbf{z}}_0) = \frac{\sqrt{\bar{\alpha}_{t-1}} \beta_t}{1 - \bar{\alpha}_t} \textit{\textbf{z}}_0 + \frac{\sqrt{\alpha_t} (1 - \bar{\alpha}_{t-1})}{1 - \bar{\alpha}_t} \textit{\textbf{z}}_t$, $\tilde{\beta}_t = \frac{1 - \bar{\alpha}_{t-1}}{1 - \bar{\alpha}_t} \beta_t$, and $\bar{\alpha}_t = \prod_{s=1}^t \alpha_s$. 
Equivalently, we have
\begin{align} 
\textit{\textbf{z}}_{t-1} &= \tilde{\mu}_t(\textit{\textbf{z}}_t, \textit{\textbf{z}}_0) + \sqrt{\tilde{\beta}_t} \boldsymbol{\epsilon}.
\end{align}
Then we can sample $\textit{\textbf{z}}_0$ from $\textit{\textbf{z}}_T$ by recursively applying
\begin{align} 
\label{eq:sampling}
\textit{\textbf{z}}_{t-1} &= \tilde{\mu}_t(\textit{\textbf{z}}_t, g_{\btheta}(\textit{\textbf{z}}_t, t)) + \sqrt{\tilde{\beta}_t} \boldsymbol{\epsilon}.
\end{align}

\section{Methodology}
\label{sec:method}

We here propose our LDM method tailored for conditional reservoir facies generation with maximal preservation of conditional data.
We first describe the differences between image generation and reservoir facies generation that are important to be considered in the design of our method,  and then outline the proposed method.

\subsection{Differences between Image Generation and Reservoir Facies Generation}
\label{sect:Differences between Image Generation and Reservoir Facies Generation}

There are several distinct differences between the image generation problem and the reservoir facies generation problem that pose challenges in employing an LDM for reservoir facies generation:

\paragraph{Input Types}
In image generation, an input image is considered continuous and one has $\textit{\textbf{x}} \in \mathbb{R}^{3 \times H \times W}$ where 3, $H$, and $W$ denote RGB channels, height, and width, respectively. In the reservoir facies generation, however, the input is categorical $\textit{\textbf{x}} \in \mathbb{Z}_2^{F \times H \times W}$ where $\mathbb{Z}_2 \in \{0, 1\}$, $F$ denotes the number of facies types, and each pixel, denoted by $\textit{\textbf{x}}_{:hw} = \textit{\textbf{x}}_{fhw} \forall
f=1,2,\dots,F$, is a one-hot-encoded vector where $1$ for a corresponding facies type index, $0$ otherwise. The conditional data of $\textit{\textbf{x}}$, notated as $\textit{\textbf{x}}_c$ has a dimension of $(F+1 \times H \times W)$. It has one more dimension than $\textit{\textbf{x}}$ for indicating a masked region.

\paragraph{Properties of Conditional Data}
The domains of conditional data in LDM are often different from the target domain. It can for instance be a text prompt, which is among the most common conditional domains. In the conditional reservoir facies generation, unlike common applications of LDMs, the conditional domain corresponds to the target domain. Importantly, its conditional data $\textit{\textbf{x}}_c$ is spatially aligned with $\textit{\textbf{x}}$. 

\paragraph{Strict Requirement to Preserve Conditional Data in Generated Sample}
In an LDM, the conditioning process has cross-attention \citep{vaswani2017attention} between the intermediate representations of the U-Net and the representation of conditional data obtained with $\mathcal{E}_c$. 
% It can be viewed that the conditional information is mapped to the U-Net as auxiliary information given the conditional generation process $p_\theta(z_{0:T} | x_c)$.
One way of viewing this is that the encoded conditional data is mapped to the U-Net as auxiliary information.
However, an LDM has a caveat in the conditional generation -- that is, its conditioning mechanism is not explicit but rather implicit. 
To be more specific, conditional data is provided to the denoising autoencoder, but the denoising process is not penalized for insufficiently honoring the conditional data.
% there is no explicit modeling on preserving the conditional information in the generated sample. 
% The model implicitly learns the conditional distribution by minimizing Eq.~\eqref{eq:L_LDM_c} or Eq.~\eqref{eq:L_LDM}. 
As a result, LDMs are often unable to fully preserve conditional data in the generated sample but rather only capture the context of conditional data. The limitation has been somewhat alleviated using classifier-free guidance \citep{ho2022classifier}, but the problem still persists. 
% Fig.~{} shows examples of generated samples by LDM, where the above mentioned limitation is visible.  
Our problem with facies generation requires precise and strict preservation of conditional data in the generated data. In other words, facies measurements in wells should be retained in the predicted facies realization. Therefore, an explicit conditioning mechanism needs to be incorporated.

\subsection{Proposed Method}
\begin{figure}
  \centering
  \includegraphics[width=0.99\textwidth]{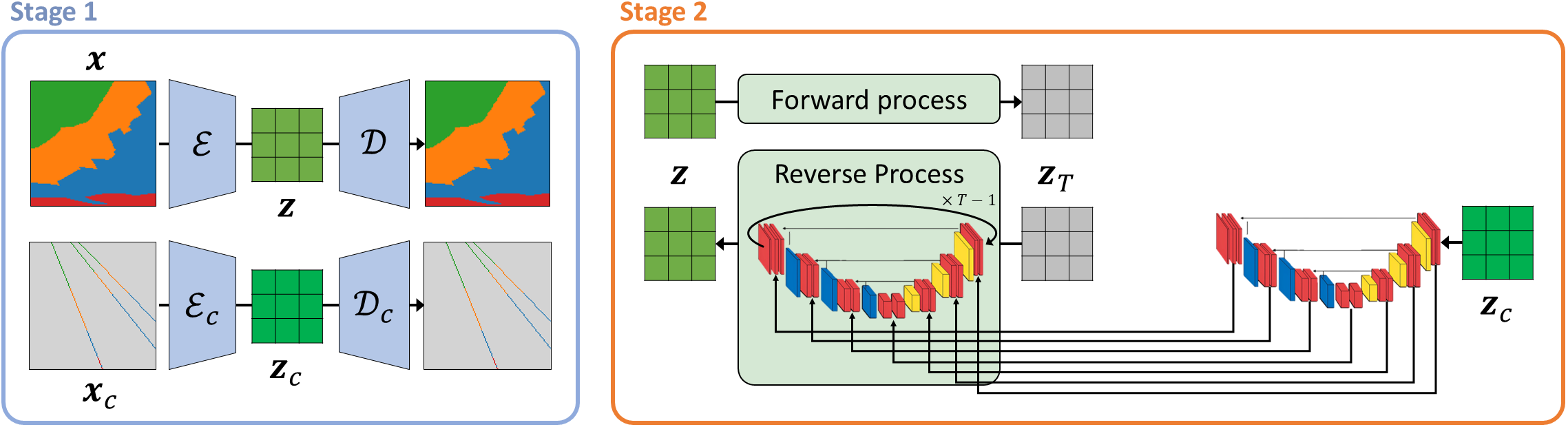}
  \caption{Overview of the training process of our proposed method. 
  It consists of two subsequent training stages: stage~1 for learning to compress and decompress data, and stage~2 for learning the conditional denoising process.
  Two U-Nets are used in stage~2. One is for the denoising process and the other is for extracting the intermediate representations of data latent vector $\textit{\textbf{z}}_c$.
  After completing the training process, the generation of a new sample (sampling process) involves the denoising of variable latent vector $\textit{\textbf{z}}_T$ into $\textit{\textbf{z}}_0 = \textit{\textbf{z}}$, using Eq.~\eqref{eq:sampling}, followed by its decoding into the data space, which is expressed as $\mathcal{D}(\textit{\textbf{z}})$.
  % In stage~2, the intermediate representations of $z_c$ obtained with the U-Net are mapped onto the re
  }
  \label{fig:overview_our_method}
\end{figure}

Our proposed method, tailored for conditional reservoir facies generation, is based on LDMs, leveraging its computational efficiency and resulting feasibility. 
The suggested method addresses several key aspects, including proper handling of the categorical input type, effective mapping of conditional data to the generative model, and maximal preservation of conditional data through a dedicated loss term for data preservation. 
Fig.~\ref{fig:overview_our_method} presents the overview of the training process of our proposed method.

\paragraph{Stage~1}
has two pairs of encoder and decoder, trained to compress and decompress $\textit{\textbf{x}}$ and $\textit{\textbf{x}}_c$, respectively. The first pair is $\mathcal{E}$ and $\mathcal{D}$ for the unconditional part, and the second pair is $\mathcal{E}_c$ and $\mathcal{D}_c$ for the conditional part. 
Here, $\mathcal{E}$ compresses $\textit{\textbf{x}}$ to $\textit{\textbf{z}}$, while $\mathcal{E}_c$ compresses $\textit{\textbf{x}}_c$ to $\textit{\textbf{z}}_c$. 
Because $\textit{\textbf{x}}$ and $\textit{\textbf{x}}_c$ are spatially aligned, we use the same architectures for $\mathcal{E}$ and $\mathcal{E}_c$, and $\mathcal{D}$ and $\mathcal{D}_c$.
Furthermore, our input is categorical as $\textit{\textbf{x}} \in \mathbb{Z}_2^{F \times H \times W}$ and $\textit{\textbf{x}}_c \in \mathbb{Z}_2^{F+1 \times H \times W}$. Therefore, we cannot naively use the stage~1 loss of LDM in Eq.~\eqref{eq:LDM_stage1_loss}. We tackle the limitation by reformulating the task as a classification task instead of a regression.
Hence, our loss function in stage~1 is
based on the cross-entropy loss function and it is formulated as:
\begin{subequations}
\begin{align} 
\label{eq:L_stage1}
L_{\mathrm{stage1}} &= \mathbb{E}_{\textit{\textbf{x}},h,w} \left[ -\sum_f \textit{\textbf{x}}_{fhw} \log{\mathrm{softmax}(\mathcal{D}(\mathcal{E}(\textit{\textbf{x}}))_{fhw})} -\sum_f (\textit{\textbf{x}}_c)_{fhw} \log{\mathrm{softmax}(\mathcal{D}_c(\mathcal{E}_c(\textit{\textbf{x}}_c))_{fhw})} \right] \\
&= \mathbb{E}_{\textit{\textbf{x}},h,w} \left[ -\sum_f \textit{\textbf{x}}_{fhw} \log \tilde{\textit{\textbf{x}}}_{fhw} -\sum_f (\textit{\textbf{x}}_c)_{fhw} \log{(\tilde{\textit{\textbf{x}}}_c)_{fhw}} \right] \\
&= CE \left( \textit{\textbf{x}}, \tilde{\textit{\textbf{x}}} \right) + CE\left(\textit{\textbf{x}}_c, \tilde{\textit{\textbf{x}}}_c \right) \\
&= L_{\mathrm{recons}} \left( \textit{\textbf{x}},\mathcal{E},\mathcal{D} \right) + L_{\mathrm{recons}}\ \left( \textit{\textbf{x}}_c,\mathcal{E}_c,\mathcal{D}_c \right), \label{eq:L_recons}
\end{align}
\end{subequations}
where $CE$ denotes a cross-entropy loss function and $L_{\text{recons}}$ denotes a reconstruction loss function.

\paragraph{Stage~2} is dedicated to learning prior and posterior distributions via learning the reverse denoising process.
The learning process involves two important perspectives: 
1) effective mapping of $\textit{\textbf{z}}_c$ to the denoising autoencoder $g_{\btheta}$ to enable the conditional generation and
2) maximal preservation of conditional data in the generated data.

To achieve the effective mapping of $\textit{\textbf{z}}_c$ to $g_{\btheta}$, we employ two U-Nets with the same architecture to process $\textit{\textbf{z}}$ and $\textit{\textbf{z}}_c$, respectively. The first U-Net is the denoising autoencoder $g_{\btheta}$ and the second U-Net is denoted $g_{\bphi}$ for extracting multi-level intermediate representations of $\textit{\textbf{z}}_c$. 
In the conditional denoising process, the intermediate representations of $\textit{\textbf{z}}_c$ are mapped onto those of $\textit{\textbf{z}}_t$ obtained with $g_{\btheta}$. This multi-level mapping enables a more effective conveyance of $\textit{\textbf{z}}_c$ which in turn results in better preservation of conditional data in the generated facies realizations.
The multi-level mapping is possible because $\textit{\textbf{x}}$ and $\textit{\textbf{x}}_c$ are spatially aligned, and equivalently for $\textit{\textbf{z}}$ and $\textit{\textbf{z}}_c$ with their intermediate representations from the U-Nets.

To achieve maximal preservation of conditional data, we explicitly tell the generative model to preserve $\textit{\textbf{x}}_c$ in the generated sample by introducing the following loss:
% \begin{subequations}
\begin{align} 
\label{eq:L_preserv}
L_{\mathrm{preserv}} &= CE\left(\textit{\textbf{x}}_c, \hat{\textit{\textbf{x}}}_c\right), 
\end{align}
% \end{subequations}
where $\hat{\textit{\textbf{x}}}_c$ represents a softmax prediction of $\textit{\textbf{x}}_c$ and is a subset of $\hat{\textit{\textbf{x}}}$ in which $\hat{\textit{\textbf{x}}} = \text{softmax}(\mathcal{D}(\hat{\textit{\textbf{z}}}))$ and $\hat{\textit{\textbf{z}}} \sim p_{\btheta}(\textit{\textbf{z}}|\textit{\textbf{z}}_t, g_{\bphi}(\textit{\textbf{z}}_c))$. Here, $p_{\btheta}(\textit{\textbf{z}}|\textit{\textbf{z}}_t, g_{\bphi}(\textit{\textbf{z}}_c))$ denotes the conditional probabilistic generative denoising process to sample $\hat{\textit{\textbf{z}}}$, given $\textit{\textbf{z}}_t$ and $g_{\bphi}(\textit{\textbf{z}}_c)$. % maybe it needs to be specified above; but we'll see. 

% The backpropagation of $L_\text{preserv}$, however, encounters a bottleneck problem in the conventional setting of stage~2 where an encoder and decoder are set to be untrainable (frozen). 
% This backpropagation traverses through $\mathcal{D}$, extending to both $g_{\btheta}$ and $g_{\bphi}$, and therefore $\mathcal{D}$ emerges as a problem when the parameters of $\mathcal{D}$ are fixed.
% To remove the bottleneck, we set $\mathcal{D}$ to be trainable, while $\mathcal{E}$ is still set to be untrainable (frozen) in stage~2. 

Finally, our loss function in stage~2 is defined by
\begin{align} 
\label{eq:L_stage2}
L_\mathrm{stage2} = 
\left\{ p_{\mathrm{uncond}} L_\mathrm{LDM}\left( \textit{\textbf{z}}, g_{\btheta} \right) + (1 - p_\text{uncond}) L_\mathrm{LDM,c}\left( \textit{\textbf{z}}, g_{\bphi}(\textit{\textbf{z}}_c), g_{\btheta} \right) \right\}
+ L_\text{preserv},
\end{align}
where $p_\mathrm{uncond}$ is a constant probability of unconditional generation, typically assigned a value of either 0.1 or 0.2 \citep{ho2022classifier}.

\section{Experiments}
\label{sec:results}

Our dataset comprises 5,000 synthetic reservoir facies samples. The generating facies model is motivated by data from shoreface deposits in wave-dominated shallow-marine depositional environments. For details about the geological modeling assumptions about bedset stacking and facies sampling, see Appendix~\ref{appendix:dataset}. The data samples are partitioned into training (80\%) and test datasets (20\%).  
In our experiments, we assess the effectiveness of our proposed version of an LDM for both conditional and unconditional facies generation. Furthermore, we present a comprehensive comparative analysis of our diffusion model against a GAN-based approach. Specifically, we adopt the U-Net GAN from \cite{zhang2021u} due to its similar conditional setup to ours and because it has shown good performance in terms of fidelity and sample diversity in the conditional generation of binary facies. 
% Additionally, its conditional setup closely aligns with ours.
For the details of our diffusion model and U-Net GAN, see Appendix~\ref{sect:implementation_detail_ldm} and Appendix~\ref{sect:implementation_detail_gan}, respectively.

\subsection{Conditional Facies Generation by the Proposed Diffusion Model}
% example of the denoising process is depicted in Fig.~\ref{fig:ldm_denoising_example}
    % | D(z_T) | D(z_{T-1}) | D(z_{T-2}) | ... | D(z_0) |
    % | error.map | error.map | ...

\begin{figure}
  \centering
  \includegraphics[width=0.8\textwidth]{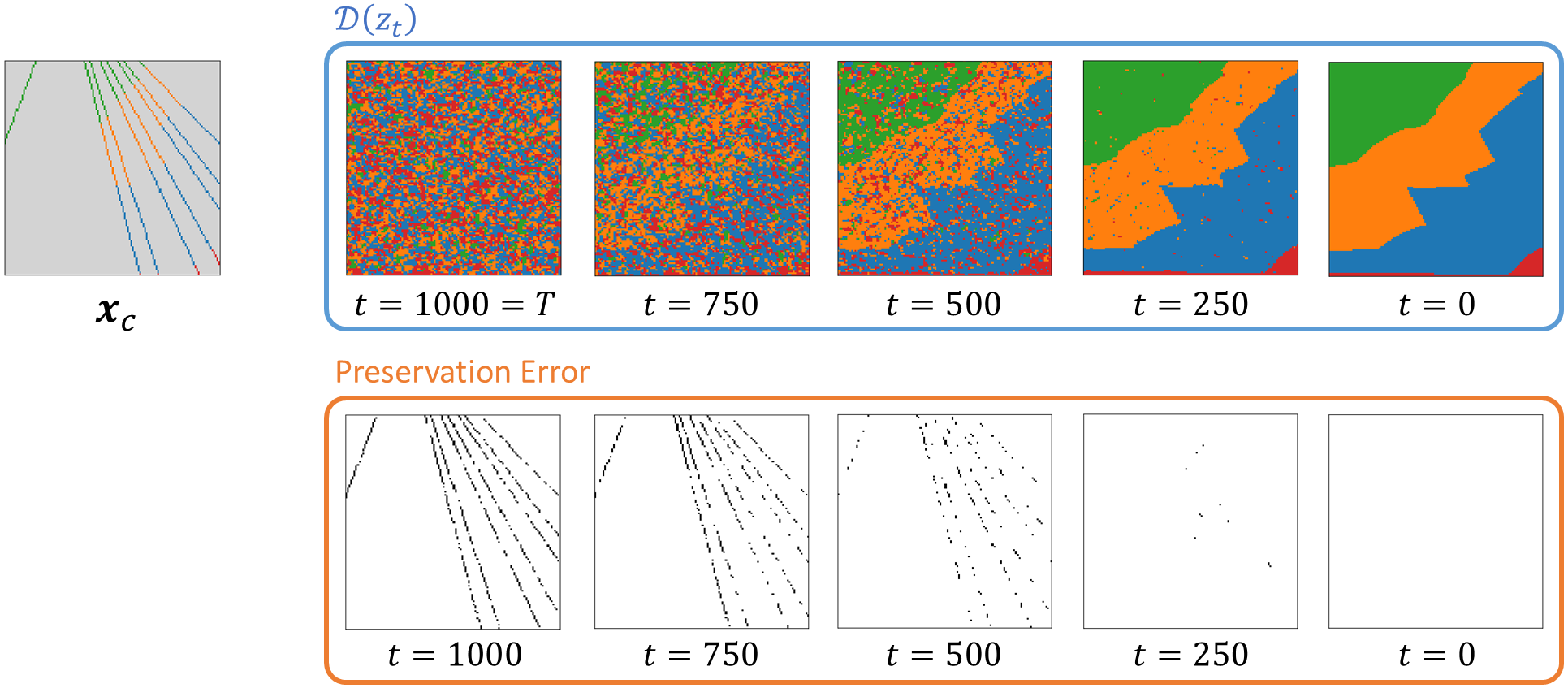}
  \caption{Visualization of transitions in the conditional denoising process.
  The denoising autoencoder sequentially denoises latent variable vector $\textit{\textbf{z}}_T$ to $\textit{\textbf{z}}_0$, conditioned on the encoded $\textit{\textbf{x}}_c$, in the sampling process. 
  Each $\textit{\textbf{z}}_t$ in the process can decoded and visualized to gain a better understanding of the conditional denoising process. 
  We present $\mathcal{D}(\textit{\textbf{z}}_t)$ at the denoising steps of 1000, 750, 500, 250, and 0, where $T=1000$ in this setup.
  The preservation error indicates the degree of accuracy with which the conditional data is retained within the generated data. Pixels colored in black indicate the error.
  }
  \label{fig:transitional_sampling_ldm}
\end{figure}

In the sampling process, the denoising autoencoder iteratively performs denoising to transition $\textit{\textbf{z}}_T$ (Gaussian noise) into $\textit{\textbf{z}}_0$ with each step being conditioned on the encoded conditional data. 
To provide a granular insight into the progressive denoising process, we present a visual example of transitions of the conditional denoising process in Fig.~\ref{fig:transitional_sampling_ldm}.
(Additional examples are presented in Fig.~\ref{fig:transitional_sampling_ldm-2} in Appendix~\ref{appendix:additional_experiments}.)
At the beginning of the denoising process ($t = 1000 = T$), $\textit{\textbf{z}}_t$ is initially composed of random Gaussian noises. Consequently, $\mathcal{D}(\textit{\textbf{z}}_t)$ also represents noise, resulting in a significant preservation error.
However, as the denoising steps progress towards $t = 0$, the generated facies gradually become more distinct and recognizable while the preservation error becomes smaller.

In Fig.~\ref{fig:transitional_sampling_ldm}, $\textit{\textbf{x}}_c$ is sourced from the test dataset, and we visualize the most probable facies types within $\mathcal{D}(\textit{\textbf{z}}_t)$. It is important to emphasize that $\mathcal{D}(\textit{\textbf{z}}_t)$ belongs to the space $\mathbb{Z}_2^{F \times H \times W}$, where the most probable facies type corresponds to the channel $f$ with the highest value.
% It is noticeable that the posterior is effectively captured by the diffusion model, producing high-fidelity reservoir facies with the robust preservation of conditional data.
The results demonstrate the effectiveness of the denoising process of our method. We notice the gradual improvement in the fidelity of the generated facies and the preservation error, eventually producing realistic facies that honor the conditional data.

\begin{figure}
  \centering
  \includegraphics[width=0.99\textwidth]{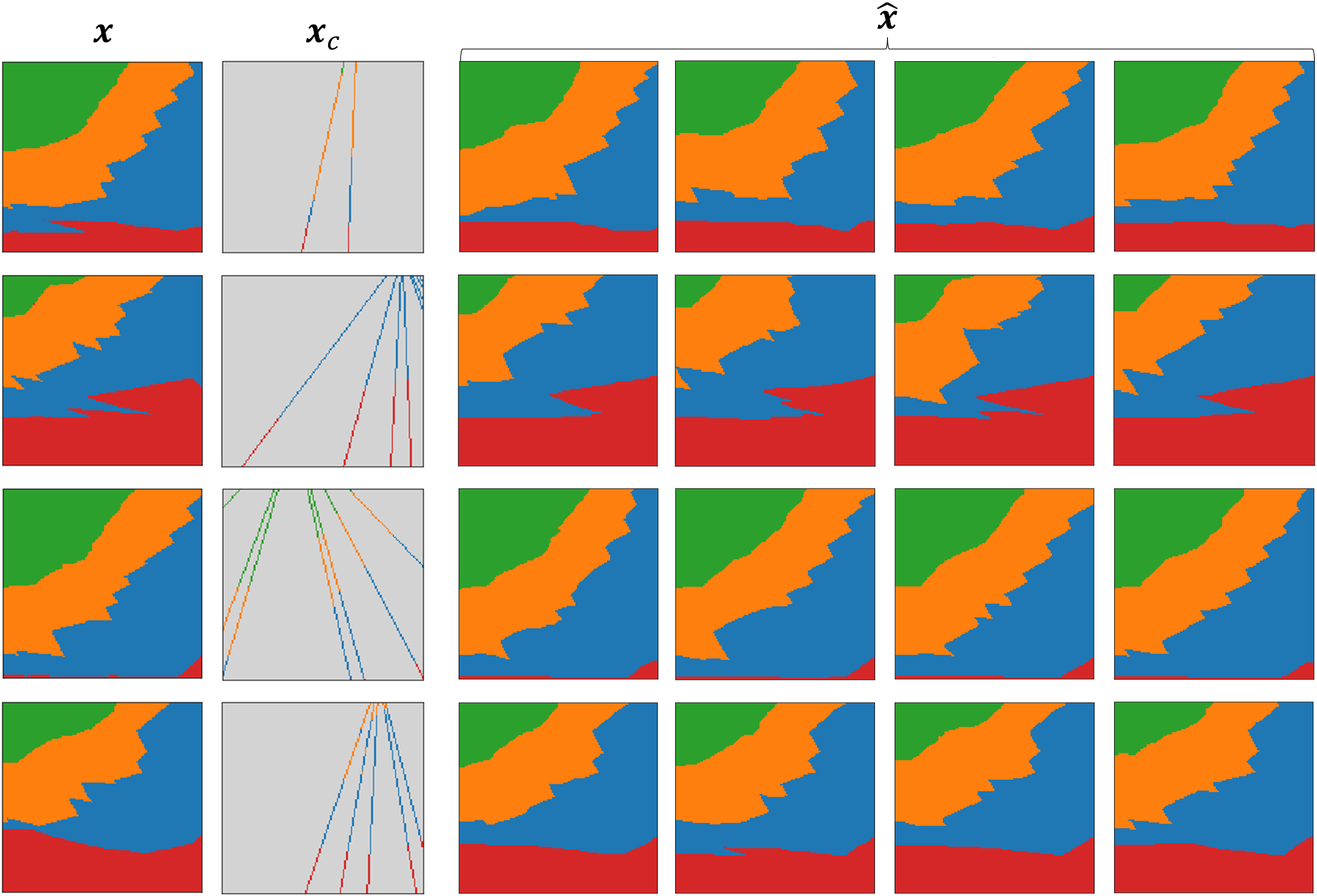} 
  \caption{Examples of multiple instances of generated facies conditioned on different conditional data $\textit{\textbf{x}}_c$ using our diffusion model. 
  The first and second columns represent $\textit{\textbf{x}}$ (ground truth) and $\textit{\textbf{x}}_c$, respectively, from the test dataset, and the remaining columns represent the conditionally generated facies.
  It is important to emphasize that the preservation error maps are omitted here because conditional sample $\hat{\textbf{x}}$ does not carry any preservation error here.
  }
  \label{fig:x_c-to-many_samples}
\end{figure}

The denoising process is stochastic, therefore various facies can be generated given $\textit{\textbf{x}}_c$. In Fig.~\ref{fig:x_c-to-many_samples}, multiple instances of conditionally-generated facies are showcased for different $\textit{\textbf{x}}_c$. Each row in this display hence represents multiple realizations of facies models, given the well facies data (second column of each row). 

The results highlight the efficacy of our diffusion model in capturing the posterior and sample diversity while adhering to given constraints. 
In particular, the conditional generation can be notably challenging, especially when there is a substantial amount of conditional data to consider (\textit{e.g.,} the third row in Fig.~\ref{fig:x_c-to-many_samples}). However, our diffusion model demonstrates its capability to honor the conditional data while generating realistic facies faithfully.
This capability facilitates the quantification of uncertainty associated with the generated facies, providing valuable insights for decision-makers in making informed decisions. With the bedset model, the well data contains much information about the transition zone from one facies type to another. This information clearly constrains the variability in the conditional samples, and there is not so much variability within the samples in a single row compared with the variability resulting from different well configurations and facies observations in the wells. 

%The sample diversity stands in contrast to other approaches such as GANs, which frequently encounter a mode collapse problem, resulting in a considerable reduction in sample diversity. This deficiency often necessitates the incorporation of an additional loss term to partially mitigate the problem \citep{zhang2021u}.

\subsection{Conditional Facies Generation by GAN and Its Limitations}
\label{sect:result_gan}

\begin{figure}
  \centering
  \includegraphics[width=.95\textwidth]{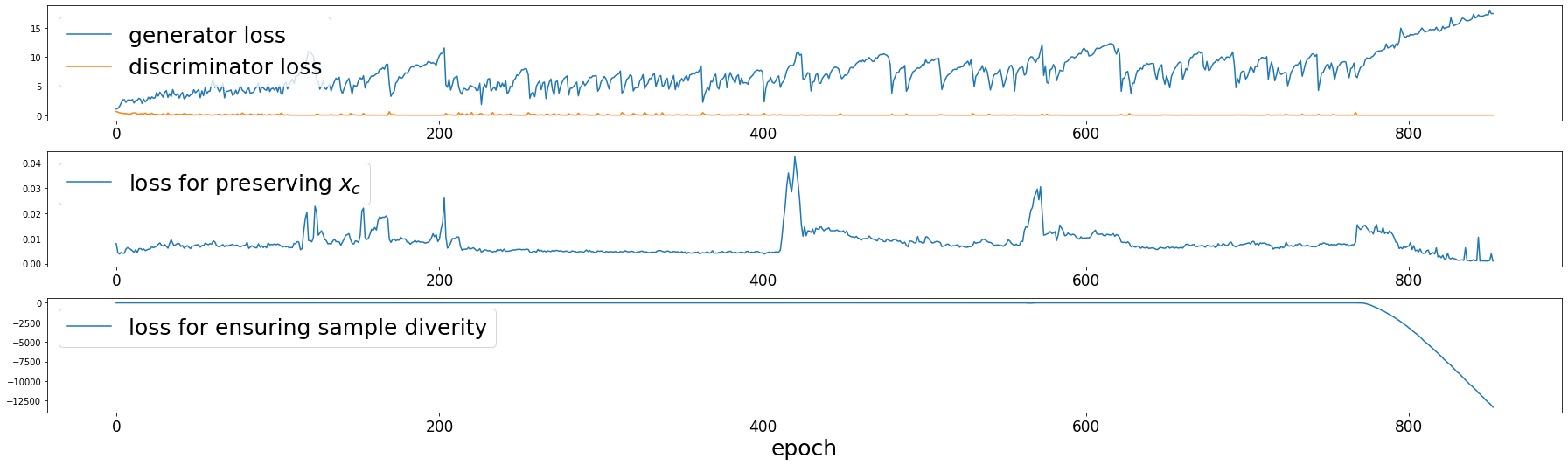}
  \caption{Training loss history of U-Net GAN. 
  The training process exhibits issues such as non-convergence, imbalance between the generator and discriminator, and divergent loss.
  }
  \label{fig:train_loss_history-gan}
\end{figure}

We next show results of using a GAN on this reservoir facies generation problem. 
As commonly observed in the GAN literature, we experienced a high level of instability in training with a U-Net GAN. 
Fig.~\ref{fig:train_loss_history-gan} presents the training history of the U-Net GAN. 
First, the gap between the generator and discriminator losses becomes larger as the training progresses. This indicates that it suffers from the generator-discriminator imbalance problem.
Second, the discriminator loss eventually converges, while the generator loss diverges towards the end of the training. This exhibits the divergent loss problem and results in a complete failure of the generator. 
Third, the loss for preserving $\textit{\textbf{x}}_c$ is unstable and non-convergent due to the unstable training process of the GAN.
Lastly, the sample diversity loss indicates better diversity when the loss value is lower and vice versa. Throughout the training process, the diversity loss remains high until shortly before around 780 epochs, then the generator fails and starts producing random images. The failure leads to a decrease in the diversity loss. It indicates that the GAN model struggles to capture a sample diversity while retaining good generative performance.

\begin{figure}
  \centering
  \includegraphics[width=\textwidth]{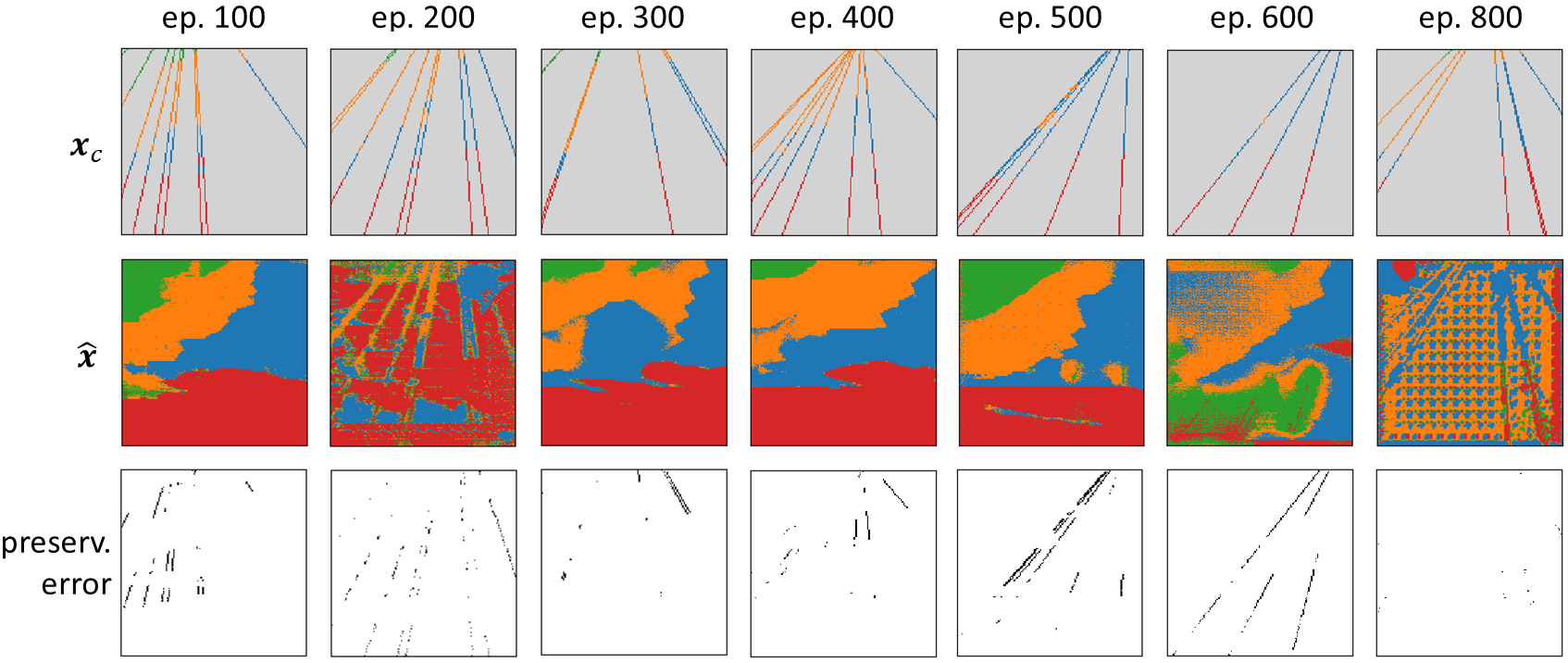}
  \caption{Visualization of conditionally-generated facies samples by U-Net GAN at different training steps with the preservation error map. Here,
  ep. denotes epoch, and variable $\textit{\textbf{x}}$ and conditioning data $\textit{\textbf{x}}_c$ are from the test dataset.
  }
  \label{fig:samples_at_different_training_steps-gan}
\end{figure}

Fig.~\ref{fig:samples_at_different_training_steps-gan} shows conditionally-generated samples using U-Net GAN at different training steps. The unstable training process can be seen in the generated samples. 
For instance, we observe a noticeable improvement in the quality of generated facies up to the 400 training epoch. 
However, from the 500 epoch, the quality continues to decline until the generated samples are barely recognizable.
Generally, the GAN model appears to face challenges in concurrently maintaining high fidelity, preserving conditional data, and achieving sample diversity, therefore failing to capture the posterior.

\begin{figure}
  \centering
  \includegraphics[width=0.99\textwidth]{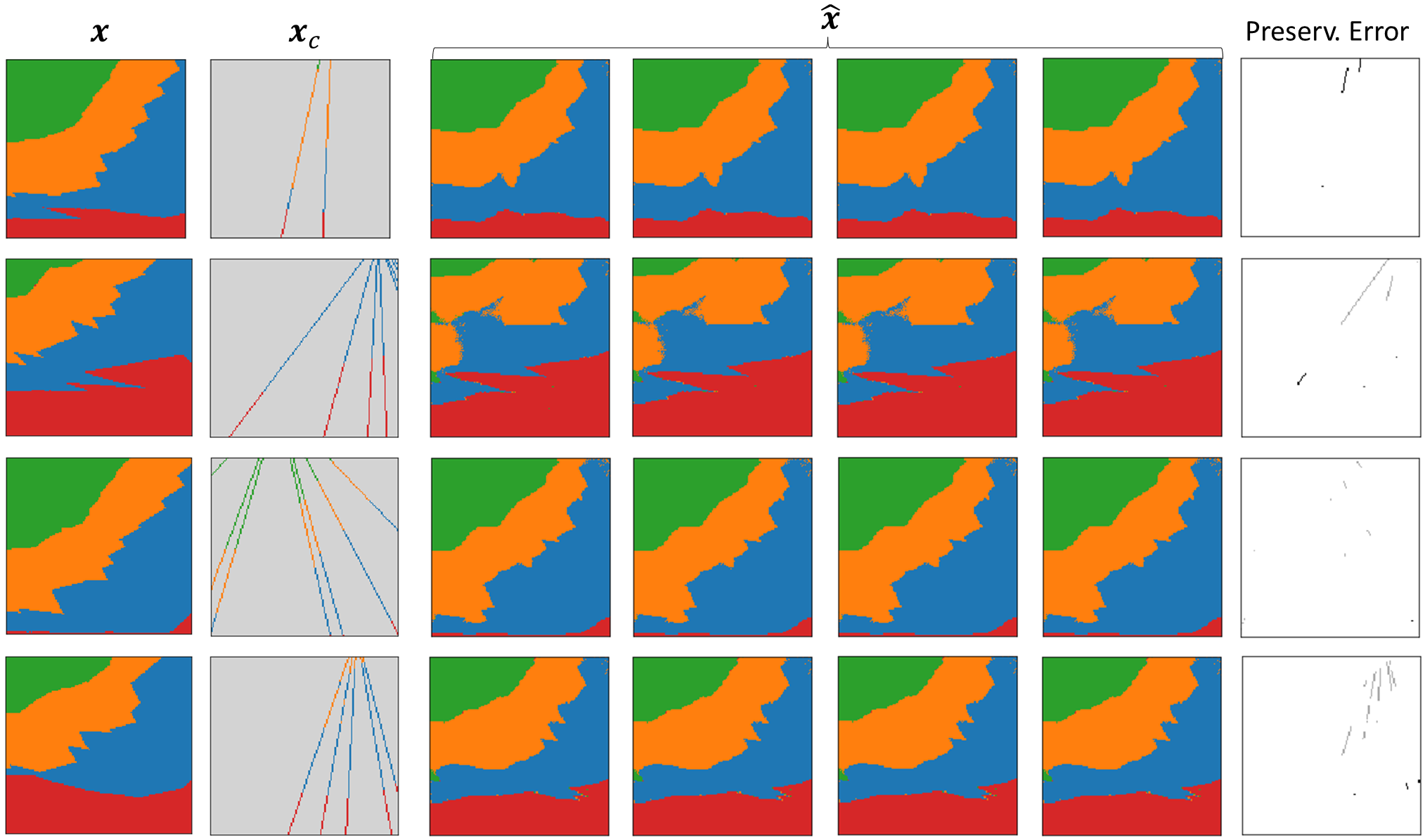}
  \caption{Examples of multiple instances of generated facies conditioned on different data $\textit{\textbf{x}}_c$ using U-Net GAN
  The first and second columns represent $\textit{\textbf{x}}$ (ground truth) and $\textit{\textbf{x}}_c$, respectively, from the test dataset, $\hat{\textit{\textbf{x}}}$ denotes the conditionally generated facies, and the last column shows the preservation error maps.
  It is important to highlight that we are showcasing a total of four distinct generated samples. Nevertheless, they appear to be identical, primarily as a result of the mode collapse phenomenon that occurs during the GAN training.
  Because the generated facies are identical, their corresponding preservation error maps are also identical. Hence, we present a single preservation error map on the right-hand side.
  }
  \label{fig:x_c-to-many_samples-gan}
\end{figure}

Fig.~\ref{fig:x_c-to-many_samples-gan} presents multiple instances of generated facies conditioned on different $\textit{\textbf{x}}_c$ using the U-Net GAN. The generator at the training epoch of 400 is used to generate the samples for its better performance than the generators at the other epochs. 
While showing more consistency than that of Fig. \ref{fig:samples_at_different_training_steps-gan}, the results still show that the generated samples have low fidelity, considerable deviations from the ground truths, and a lack of sample diversity due to the mode collapse.
Furthermore, the generated samples exhibit a considerable sum of preservation errors, indicating the incapability to retain the conditional data.
Overall, the results demonstrate that the GAN model fails to capture the posterior. 

Table~\ref{tab:preservation_err_rate} specifies the preservation error rates of our proposed diffusion model and the U-Net GAN. 
The preservation error rate is defined as 
\begin{align*}
\text{Preservation error rate} = \frac{ \text{number of different pixels between } \textit{\textbf{x}}_c \text{ and } \text{argmax } \hat{\textit{\textbf{x}}}_c \text{ for the valid pixels in } \textit{\textbf{x}}_c}{\text{number of valid pixels in } \textit{\textbf{x}}_c},
\end{align*}
where the preservation error rate of zero indicates perfect preservation.
The results demonstrate that our diffusion model achieves the near-perfect preservation (\textit{i.e.,} only 0.04\% of conditional data fails to be retained) and it significantly outperforms the U-Net GAN in retaining conditional data, surpassing it by a factor of approximately 255 times. 

\begin{table}
    \caption{Preservation error rates of our diffusion model and U-Net GAN on the test dataset. }
    \label{tab:preservation_err_rate}
    \centering
    \begin{tabular}{ccc}
        \toprule
         & Our Diffusion Model & U-Net GAN \\
         \midrule
        % Preservation error rate & \textbf{0.0005} & 0.0216 \\
        Preservation error rate & \textbf{0.0004} & 0.1022 \\
        \bottomrule
    \end{tabular}
\end{table}

To better illustrate the mode collapse phenomenon in U-Net GANs in comparison to our proposed diffusion model, Fig.~\ref{fig:prior_distributions} shows a comparison between the prior distributions of the training and test datasets along with the prior distribution predicted by our diffusion model and that of the U-Net GAN. 
To visualize the prior distributions of the training and test datasets, we first employ an argmax operation on $\textit{\textbf{x}}$ across the channel dimension. This operation results in $\text{argmax } \textit{\textbf{x}} \in \mathbb{R}^{H \times W}$ that contains integer values. Subsequently, we calculate the average of $\text{argmax } \textit{\textbf{x}}$ for all instances of $\textit{\textbf{x}}$ within the training or test dataset.
The same visualization procedure is applied to visualize the prior distributions predicted by our diffusion model and U-Net GAN, with the only difference being the application of an argmax operation to generated facies data.
For the U-Net GAN, its generator at the 400 training epochs is used (same as above).
The prior distribution contains four main distinct colors (green, orange, blue, red) depending on facies types, and darker colors indicate high likelihood and vice versa.
% The prior distribution shows four main distinct colors (dark teal, light teal, yellow, and purple), and between the different facies, there exist intermediate areas where different facies can occur, represented by the intermediate colors.
These results clearly demonstrate our diffusion model's capability to accurately capture the prior distribution, whereas the U-Net GAN faces substantial challenges in this regard due to the mode collapse, leading to severe underestimation of the variability in the generated samples.
 % which attains good sample diversity by effectively capturing the prior distribution of the target dataset

In Fig.~\ref{fig:prior_distributions} we also show the Jensen-Shannon (JS) divergence of the generated samples compared with the true model. Similar to the Kullback-Leibler divergence, but non-symmetric and finite (between $0$ and $1$), the JS divergence here measures the probabilistic difference between the generated and training samples. Clearly, the divergence is much smaller for the LDM here, while the GAN gets very large JS divergence (close to 1) at some of the facies boundaries. Even though it is less prominent than for the GAN, the divergence for LDM shows some structure near the facies transition zones. This indicates some underestimation in the implicit posterior sampling variability.

\begin{figure}
  \centering
  \includegraphics[width=0.99\textwidth]{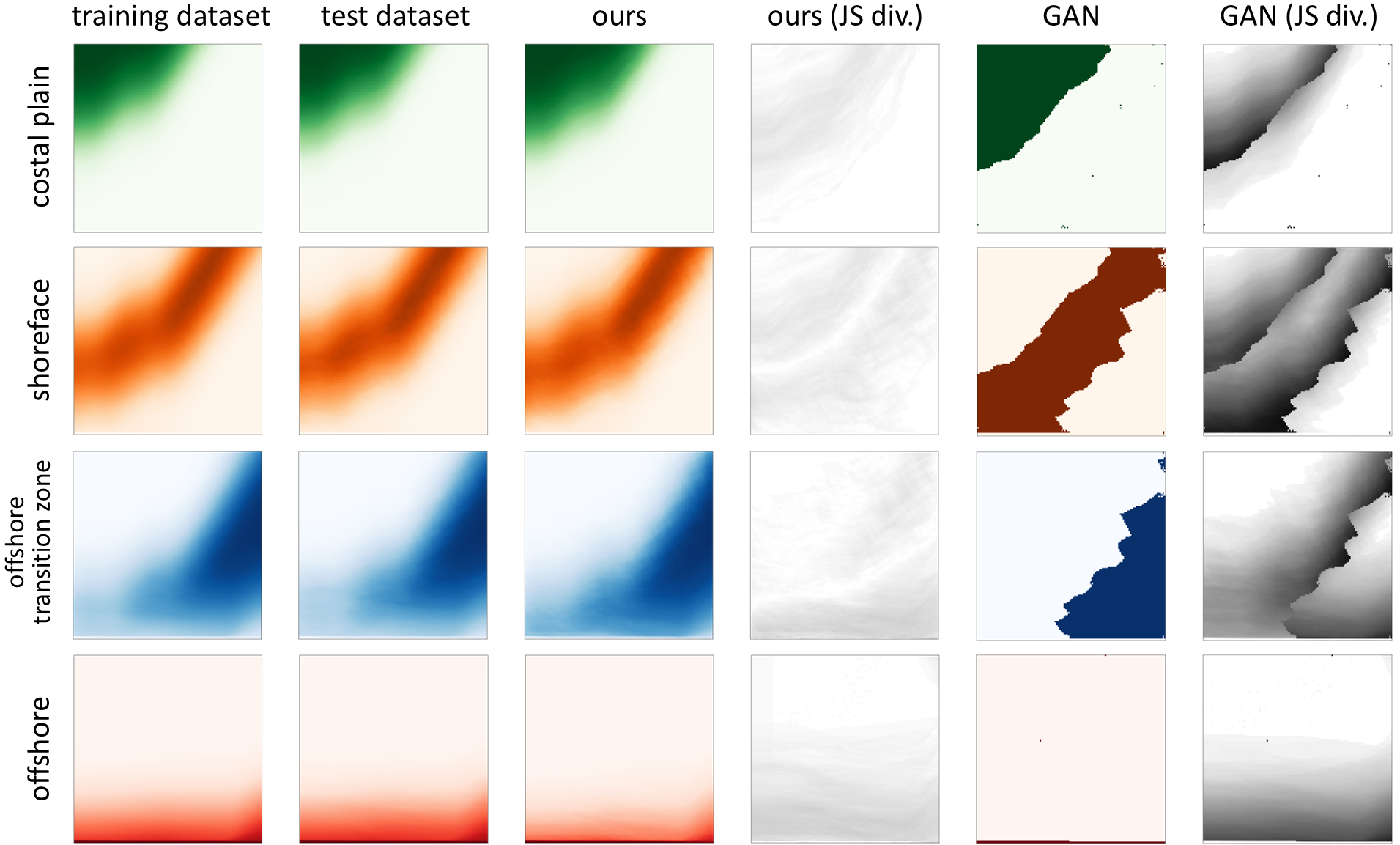}
  \caption{Comparative visualization of the prior distributions of the training and test datasets and the prior distribution predicted by our diffusion model and the U-Net GAN. 
  The binary-colored figure depicts the discrepancy, measured by JS divergence, between the prior distribution of the training dataset and the predicted prior distribution.
  At every pixel, there exists a prior distribution encompassing various facies types, where these facies types are visually represented by distinct colors. The presence of intermediate colors signifies a prior distribution with greater diversity.
  In this context, it is important to highlight that the mode collapse observed in the U-Net GAN results in a reduction of sample diversity, analogous to the absence of those intermediate colors in the visualization.
  }
  \label{fig:prior_distributions}
\end{figure}

\subsection{Ablation Study}

We conduct an ablation study to investigate the effects of the use of the proposed components such as $L_\text{preserv}$ and the multi-level mapping of $\textit{\textbf{z}}_c$. The evaluation is performed on the test dataset.
Table~\ref{tab:ablation_study_stage2_loss} outlines the specific Case (a)-(c) considered in the ablation study, and Table~\ref{tab:preservation_err_rate_ablation} reports $p_{\mathrm{uncond}} L_\mathrm{LDM} + (1 - p_\text{uncond}) L_\mathrm{LDM,c}$ from Eqs.~\eqref{eq:L_LDM-2}-\eqref{eq:L_LDM_c-2}, $L_{\text{preserv}}$, and a preservation error rate on the test set. 

When analyzing these results, several key findings emerge. Firstly, in Case (b), both $L_\text{preserv}$ and the preservation error rate exhibit a significant increase compared to Case (a). This increase can be attributed to the fact that the denoising model in (b) was not explicitly trained to preserve the conditional data, resulting in a notable degradation in preservation quality.
%
% Moving on to Case (c), we observe slightly elevated values for both $L_\text{LDM}$ and $L_{\text{LDM},c}$. This outcome can be explained by the challenges posed during training due to the ineffective backpropagation of $L_\text{preserv}$, caused by the presence of the untrainable $\mathcal{D}$ which also hinders the preservation ability. 
%
Case (c) sheds light on the effectiveness of the multi-level mapping, comparing $L_\text{preserv}$ and the preservation error rate from the baseline Case (a). It emphasizes the positive impact of the multi-level mapping approach on preservation.
Overall, the ablation study reveals that each component in our methodology plays a vital role in enhancing the overall preservation capacity of the conditional sampling.

\begin{table}
    \caption{Ablation study cases with respect to the novel and essential components in stage~2. 
    The signs of \texttt{o} and \texttt{x} indicate the use of the item described in the corresponding column name, where \texttt{o} and \texttt{x} denote using and not using, respectively. 
    In the case of (c), instead of employing multi-level mapping for $\textit{\textbf{z}}_c$, it takes a straightforward route to integrate $\textit{\textbf{x}}_c$ into the denoising U-Net. This integration is achieved through a simple concatenation of $\textit{\textbf{z}}_c$ and $\textit{\textbf{z}}_t$, forming the input denoted as $[\textit{\textbf{z}}_t, \textit{\textbf{z}}_c]$ for the denoising U-Net, where $[.]$ represents the concatenation operation.
    }
    \label{tab:ablation_study_stage2_loss}
    \centering
    \begin{tabular}{lccccc}
    \toprule
    ~ & $L_\text{preserv}$ & Multi-level mapping of $\textit{\textbf{z}}_c$ \\
    \midrule
    (a) Base & \texttt{o} & \texttt{o} \\
    (b) $-$ $L_\text{preserv}$ & \texttt{x} &  \texttt{o}  \\
    (c) $-$ Multi-level mapping of $\textit{\textbf{z}}_c$ & \texttt{o} & \texttt{x} \\ 
    \bottomrule
    \end{tabular}
\end{table}

\begin{table}
    \caption{
    Effects of the use of $L_\text{preserv}$ and multi-level mapping of $\textit{\textbf{z}}_c$ for the ablation study cases. 
    }
    \label{tab:preservation_err_rate_ablation}
    \centering
    \begin{tabular}{lccc}
        \toprule
         & (a) Base & (b) $-$ $L_\text{preserv}$ & (c) $-$ Multi-level mapping of $\textit{\textbf{z}}_c$ \\
         \midrule
         % $L_{\text{LDM}}$ & 0.0148 & 0.0138 & 0.0155 \\
         % $L_{\text{recons}}$ & 0.0026 & 0.0030 & 0.0027 \\
         $p_{\mathrm{uncond}} L_\mathrm{LDM} + (1 - p_\text{uncond}) L_\mathrm{LDM,c}$ & 0.02444 & {0.02306} & {0.02618}  \\
         $L_{\text{preserv}}$ & \textbf{0.00029} & {0.00643} & {0.00082} \\
         Preservation error rate & \textbf{0.00044} & {0.00442} & {0.00146} \\
        \bottomrule
    \end{tabular}
\end{table}

\section{Conclusion}
\label{sec:concl}

We have introduced a novel approach for conditional reservoir facies modeling employing LDM. Experimental results show exceptional abilities to preserve conditional data within generated samples while producing high-fidelity samples. 
Our novelties lie in the proposals to enhance the preservation capabilities of LDM.
Throughout our experiments, we have demonstrated the robustness and superiority of our diffusion-based method when compared to a GAN-based approach, across multiple aspects including fidelity, sample diversity, and conditional data preservation. 
Furthermore, we have presented the critical limitations of the GAN approach, which result in compromised fidelity, limited sample diversity, and sub-optimal preservation performance.

Overall, our work opens up a new avenue for conditional facies modeling through the utilization of a diffusion model. The results indicate some underestimation in the posterior samples, which can possibly improved by more nuanced training or refined loss functions. 
As future work, we aim to study the statistical properties of the LDM in detail on various geostatistical models, study conditioning to other data types, and extend our method for 3D facies modeling. 

Conditioning to seismic data can be done in various ways. For instance by introducing a seismic loss function (possibly including convolution effects in the seismic forward model) to the training loss and the reconstruction loss. 
Expanding our method for 3D facies modeling may appear straightforward by merely substituting 2D convolutional layers with their 3D counterparts. However, dealing with 3D spatial data presents inherent complexities stemming from its high-dimensional nature. This can manifest in various challenges, including high computational demands and the difficult learning of prior and posterior distributions. Therefore, it may need to employ techniques like hierarchical modeling. This can involve employing a compact latent dimension size for sampling, followed by an upscaling mechanism similar to super-resolution, in order to enhance the feasibility and effectiveness of the 3D modeling.
Moreover, incorporating additional conditional data, such as seismic information, into the sampling process can be achieved through the use of cross-attention mechanisms, as introduced in the original LDM.

\subsubsection*{Acknowledgements}
We would like to thank the Norwegian Research Council for funding the Machine Learning for Irregular Time Series (ML4ITS) project (312062), the GEOPARD project (319951) and the SFI Centre for Geophysical Forecasting (309960). 

\subsubsection*{Ethical Statement}
No conflicts of interest were present during the research process.

%Bibliography
%\clearpage
%\newpage
\bibliography{bibliography}

\newpage
\appendix

\section{Dataset}
\label{appendix:dataset}
% how the data is generated; more in detail

Our test images are vertical 2D slices through a shoreface deposit in a wave-dominated shallow-marine depositional environment. The cross section is taken along the dip direction, with the proximal or landward side to the left in the image, and the distal or seaward side to the right. The shoreface deposit consists of sediment packages referred to as bedsets. On the landward side of the bedsets is the coastal plain, which is coaly and of poor reservoir quality. The proximal part of each bedset consists of shoreface sand of good reservoir quality, while the distal part has sand interbedded with shale. The offshore region beyond the distal edge of the bedsets has only shale.

We create realizations using the rule-based object model GEOPARD, which was described by \citet{scotti2022defining}. This model sequentially stacks bedsets in a way that mimics the depositional process. Bedset-scale description is appropriate for reservoir facies modeling \citep{Bedsets}. The trajectory of the shoreline is a function of bedset progradation and aggradation, in other words how much the bedsets build out and build up. These, in turn, are controlled by such environmental factors as the sea level and sediment supply. See also the article by \citet{ovanger2024addressing}, where a conceptually similar shoreface deposition model is considered. GEOPARD first generates base and top surfaces for a sequence of bedsets, and then uses these surfaces to create a 3D grid of facies values. The data in this study consists of 2D arrays extracted from these 3D grids. We take one slice from each 3D grid. That is, we do not take multiple slices from the same realization. Facies is treated as a categorical variable and one-hot encoded, as described in Sect.~\ref{sect:Differences between Image Generation and Reservoir Facies Generation}. 

The conditional data $\textit{\textbf{x}}_c$ are generated by taking a subset of $\textit{\textbf{x}}$ with a random number of straight lines and random angles within certain ranges. The line patterns resemble groups of deviated wells with a common template, in other words originating from a common point somewhere above the image. The number of lines is sampled from a Poisson distribution with expectation four and then shifted up by one so that the expected number of lines is five, and there is always at least one line. Our dataset comprises 5000 facies realizations, split into training (80\%) and test datasets (20\%). The full dataset is available at \url{https://figshare.com/articles/dataset/Dataset_used_in_Latent_Diffusion_Model_for_Conditional_Reservoir_Facies_Generation/26892868?file=48931588}.

\section{Implementation Details of Our Proposed Method}
\label{sect:implementation_detail_ldm}

% \subsection{Models}
% models details (incl. hyperparameters): encoder, decoder, U-Net, LDM
    % decoder includes VQ
    % data scaling in encoder or VQ
        % point out the issue of non-determined data range for z unlike x.
        % In DM, clipping on x is done between -1 and 1 [REF]
        % We'd like to ensure such a range for z as well. To acheive that, we use
            % z_q = z_q/max(z_q) within the decoder
    % optimizer
\subsection{Encoders and Decoders: $\mathcal{E}$, $\mathcal{E}_c$, $\mathcal{D}$, and $\mathcal{D}_c$}
The same encoder and decoder architectures from the VQ-VAE paper are used and their implementations are from \url{https://github.com/nadavbh12/VQ-VAE}. 
% The encoder consists of $n$ downsampling convolutional blocks (Conv2d -- BatchNorm2d -- LeakyReLU), followed by $m$ residual blocks (LeakyReLU -- Conv2d -- BatchNorm2d -- LeakyReLU -- Conv2d). The short notations are taken from the PyTorch implementations.
The encoder is a stack of a downsampling convolution block (Conv2d -- BatchNorm2d -- GELU -- Dropout) and a subsequent residual block (GELU -- Conv2d -- BatchNorm2d -- GELU -- Dropout -- Conv2d). The short notations are taken from the PyTorch implementations.
The architecture of the decoder is the inverse of the encoder's. Its upsampling convolutional layer is implemented with (Upsample(mode='nearest') -- Conv2d).
In the encoding process, the spatial size halves and the hidden dimension doubles after every downsampling block, and the bottleneck dimension (\textit{i.e.,} dimension of $\textit{\textbf{z}}$ and $\textit{\textbf{z}}_c$) is set to 4, following \cite{rombach2022high}.

The stack size determines a downsampling rate. For instance, a single stack corresponds to a downsampling rate of 2. In our experiments, we use a single stack because we observed that a higher downsampling rate leads to higher loss of input information, resulting in an inadequate reconstruction of $\textit{\textbf{x}}_c$. This inadequacy suggests that $\textit{\textbf{z}}_c$ fails to fully capture the information contained in $\textit{\textbf{x}}_c$, ultimately leading to a deficiency in preserving $\textit{\textbf{x}}_c$ within a generated sample. In addition, \cite{rombach2022high} demonstrated that a low compression rate is sufficient for LDM to generate high-fidelity samples.

In the naive form of $\mathcal{E}$ and $\mathcal{E}_c$, the value ranges of $\textit{\textbf{z}}$ and $\textit{\textbf{z}}_c$ are not constrained. However, the diffusion model $g_{\btheta}$ is typically designed to receive a value ranging between -1 and 1. For instance, image data is scaled to range between -1 and 1 to be used as the input. 
To make $\textit{\textbf{z}}$ and $\textit{\textbf{z}}_c$ compatible with the diffusion model, we normalize them as $\textit{\textbf{z}} / \max(|\textit{\textbf{z}}|)$ and $\textit{\textbf{z}}_c / \max(|\textit{\textbf{z}}_c|)$, respectively. 

% Furthermore, we enhance the model by incorporating learnable positional embeddings, as demonstrated in \cite{dosovitskiy2020image}, which are added to both $\textit{\textbf{z}}$ and $\textit{\textbf{z}}_c$ through concatenation. This addition is motivated by the presence of specific positional patterns within the facies data, and these learnable positional embeddings have the capability to capture and leverage such structures, introducing an inductive bias that helps the model generalize better to the typical positions of facies. The effectiveness of this approach was demonstrated in \cite{dosovitskiy2020image}.

% Lastly, our LDM adopts \textit{VQ-reg} for $\mathcal{D}$ and $\mathcal{D}_c$ \citep{rombach2022high}. VQ-reg denotes that a vector quantization layer \citep{van2017neural} is included in the decoder. To be more specific, the decoding process involves quantizing $\textit{\textbf{z}}$ first and decoding it to $\textit{\textbf{x}}$. 

\subsection{U-Net}
% two unets are used in our proposed method.
% the same architectures to have the same spatial dimensions.
% we use the UNet implementation from []
% parameter setting description
Two U-Nets are used in our proposed method -- one for $g_{\btheta}$ and the other for processing $\textit{\textbf{z}}_c$. The two U-Nets have the same architecture to have the same spatial dimensions for the multi-level mapping. 
% We use the implementation of U-Net from \url{https://github.com/lucidrains/denoising-diffusion-pytorch}. 
We use the implementation of U-Net from here\footnote{\url{https://github.com/lucidrains/denoising-diffusion-pytorch}}. 
Its default parameter settings are used in our experiments except for the input channel size and hidden dimension size. 
To be more precise, we use \texttt{in\_channels} (input channel size) of 4 because it is the dimension sizes of $\textit{\textbf{z}}$ and $\textit{\textbf{z}}_c$) and \texttt{dim} (hidden dimension size) of 64.
% , and \texttt{dim\_mults} of (1, 2, 4, 8).
% In our experiment, the input channel and hidden dimension sizes are set to 4 (\textit{i.e.,} same as the dimensions of $\textit{\textbf{z}}$ and $\textit{\textbf{z}}_c$) and 64, respectively. 

\subsection{Latent Diffusion Model}
% LDM: E,D + DDPM
% we use the DDP implementation from []
% parameter setting description
LDM is basically a combination of the encoders, decoders, and DDPM, where DDPM is present in the latent space.
We use the implementation of DDPM from here\footnote{See footnote 1}. %\cite{ddpm_github}.
Its default parameter settings are used in our experiments except for the input size and denoising objective for which we use the prediction of $\textit{\textbf{z}}_0$ instead of $\boldsymbol{\epsilon}$, as described in Sect.~\ref{sect:Latent Diffusion Model}.

\subsection{Optimizer}
% AdamW
% its parameter settings
We employ the Adam optimizer \citep{kingma2014adam}. We configure batch sizes of 64 and 16 for stage~1 and stage~2, respectively. The training periods are 100 epochs for stage~1 and 20000 steps for stage~2.

\subsection{Unconditional Sampling}
% unconditional sampling
    % [explain how to do that: replacing the representations of z_c with mask vectors, I think]
% [maybe] classifier-free guidance
The conditional sampling is straightforward as illustrated in Fig.~\ref{fig:overview_our_method}. For the unconditional sampling, we replace $\textit{\textbf{z}}_c$ with mask tokens, typically denoted as $\texttt{[MASK]}$ or $\texttt{[M]}$ \citep{lee2023vector,lee2024explainable}. The role of the mask token is to indicate that the sampling process is unconditional. The mask token is a learnable vector trained in stage~2 by minimizing $L_\text{LDM}(\textit{\textbf{z}},g_\theta)$ in $L_\text{stage2}$.

\section{Implementation Details of U-Net GAN}
\label{sect:implementation_detail_gan}
We implement U-Net GAN, following the approach outlined in its original paper \citep{zhang2021u}. Two key hyperparameters govern the weighting of loss terms in this implementation: one for preserving conditional data (content loss) and the other for ensuring sample diversity (diverse loss). We maintain the same weights as specified in the paper, with a value of 0.05 for the diverse loss and 100 for the content loss.
For optimization, we employ the Adam optimizer with a batch size of 32, a maximum of 750 epochs, and a learning rate set to 0.0002.
The implementation is included in our GitHub repository.

To enhance the training of GANs, several techniques like feature matching, historical averaging, and one-sided label smoothing have been proposed. These methods, detailed in \cite{salimans2016improved}, are primarily aimed at stabilizing the training process. However, in our approach, we have chosen not to implement these techniques. Instead, we focus on maintaining the fundamental structure of the GAN model to assess its performance in a basic form.

\section{Pseudocode}
To increase the reproducibility of our work and understanding of our codes in our GitHub repository, we present a pseudocode of the training process of our method in Algorithm~\ref{alg:pseudocode_stage1} for stage~1 and Algorithm~\ref{alg:pseudocode_stage2} for stage~2.
In the pseudocodes, we provide a more detailed specification of $\mathcal{D}$. 

\begin{algorithm}
\caption{Pseudocode of the training process of the proposed diffusion model (stage~1)}
\label{alg:pseudocode_stage1}
\begin{algorithmic}
\vspace{0.2em}

\While{a maximum epoch is not reached}
% \For{$\textit{\textbf{x}} \in \textit{\textbf{X}}$} 
    \State sample $\textit{\textbf{x}}$ from $\textit{\textbf{X}}$  \Comment{$\textit{\textbf{X}}$ denotes a training dataset. In practice, a batch of $\textit{\textbf{x}}$ is sampled.}
    \State $\textit{\textbf{x}}_c \gets \text{stochastically extracting conditional well data from } \textit{\textbf{x}}$ 
    \\
    \State $\textit{\textbf{z}}, \textit{\textbf{z}}_c \gets \mathcal{E}(\textit{\textbf{x}}), \mathcal{E}_c(\textit{\textbf{x}}_c)$
    % \State $\textit{\textbf{z}}_q, \ell_{VQ} = VQ(\textit{\textbf{z}})$  \Comment{$\ell_{VQ}$ denotes a loss from VQ to reduce a gap between $\textit{\textbf{z}}$ and $\textit{\textbf{z}}_q$ \citep{van2017neural}.}
    % \State $(\textit{\textbf{z}}_q)_c, \ell_{VQ_c} = VQ_c(\textit{\textbf{z}}_c)$

    \State $\tilde{\textit{\textbf{x}}}, \tilde{\textit{\textbf{x}}}_c \gets \text{softmax}( \mathcal{D}(\textit{\textbf{z}})), \text{softmax}( \mathcal{D}_c(\textit{\textbf{z}}_c))$
    \\
    \State $L_{\text{stage1}} \gets CE(\textit{\textbf{x}}, \tilde{\textit{\textbf{x}}}) + CE(\textit{\textbf{x}}_c, \tilde{\textit{\textbf{x}}}_c)$
    \\
    \State update $\mathcal{E}$, $\mathcal{E}_c$, $\mathcal{D}$, and $\mathcal{D}_c$ by minimizing $L_\text{stage1}$
% \EndFor
\EndWhile

\end{algorithmic}
\end{algorithm}

\begin{algorithm}
\caption{Pseudocode of the training process of the proposed diffusion model (stage~2)}
\label{alg:pseudocode_stage2}
\begin{algorithmic}
\vspace{0.2em}

\State load the pretrained $\mathcal{E}$, $\mathcal{E}_c$, $\mathcal{D}$, and $\mathcal{D}_c$ and freeze them.
\State randomly initialize $g_{\btheta}$ and $g_{\bphi}$

\While{a maximum epoch is not reached}
% \For{$\textit{\textbf{x}} \in \textit{\textbf{X}}$} 
    \State sample $\textit{\textbf{x}}$ from $\textit{\textbf{X}}$
    \State $\textit{\textbf{x}}_c \gets \text{stochastically extracting conditional well data from } \textit{\textbf{x}}$ 
    \\
    \State $\textit{\textbf{z}}, \textit{\textbf{z}}_c \gets \mathcal{E}(\textit{\textbf{x}}), \mathcal{E}_c(\textit{\textbf{x}}_c)$  \Comment{$\textit{\textbf{z}} = \textit{\textbf{z}}_0$}
    \State $\textit{\textbf{z}}_t \gets \text{forward diffusion process applied to } \textit{\textbf{z}}_0$  \Comment{adding noise to $\textit{\textbf{z}}_0$}

    \If {$r \leq p_\text{uncond}$}  \Comment{$r \sim U(0, 1)$ where $U$ denotes a uniform distribution}
        \State $\hat{\textit{\textbf{z}}}_0 \gets g_{\btheta}(\textit{\textbf{z}}_t, t)$  \Comment{unconditional generation}
        \State $L_{\text{LDM}} \gets \| \textit{\textbf{z}}_0 - \hat{\textit{\textbf{z}}}_0 \|_2^2$
        \State $\ell_\text{LDM} \gets L_{\text{LDM}}$
    \Else 
        \State $\hat{\textit{\textbf{z}}}_0 \gets g_{\btheta}(\textit{\textbf{z}}_t, t, g_{\bphi}(\textit{\textbf{z}}_c))$  \Comment{conditional generation}
        \State $L_{\text{LDM},c} \gets \| \textit{\textbf{z}}_0 - \hat{\textit{\textbf{z}}}_0 \|_2^2$
        \State $\ell_\text{LDM} \gets L_{\text{LDM},c}$
    \EndIf
    \\
    % \State $\hat{\textit{\textbf{z}}}_q \gets VQ(\hat{\textit{\textbf{z}}})$  \Comment{$\hat{\textit{\textbf{z}}} = \hat{\textit{\textbf{z}}}_0$}
    \State $\hat{\textit{\textbf{x}}} = \text{softmax}(\mathcal{D}(\hat{\textit{\textbf{z}}}))$  \Comment{$\hat{\textit{\textbf{z}}} = \hat{\textit{\textbf{z}}}_0$}
    \State $\hat{\textit{\textbf{x}}}_c \gets$ retrieving the valid pixel locations in $\textit{\textbf{x}}_c$ from $\hat{\textit{\textbf{x}}}$
    \State $L_\text{preserv} \gets CE(\textit{\textbf{x}}_c, \hat{\textit{\textbf{x}}}_c)$
    \\
    \State $L_\text{stage2} \gets \ell_\text{LDM} + L_\text{preserv}$
    \\
    \State update $g_{\btheta}$ and $g_{\bphi}$ by minimizing $L_\text{stage2}$
    
% \EndFor
\EndWhile

\end{algorithmic}
\end{algorithm}

\section{Additional Experimental Results}
\label{appendix:additional_experiments}

In continuation of Fig.~\ref{fig:transitional_sampling_ldm}, additional examples of the transitions in the conditional denoising process are presented in Fig.~\ref{fig:transitional_sampling_ldm-2}.

\begin{figure}
  \centering
  \includegraphics[width=0.65\textwidth]{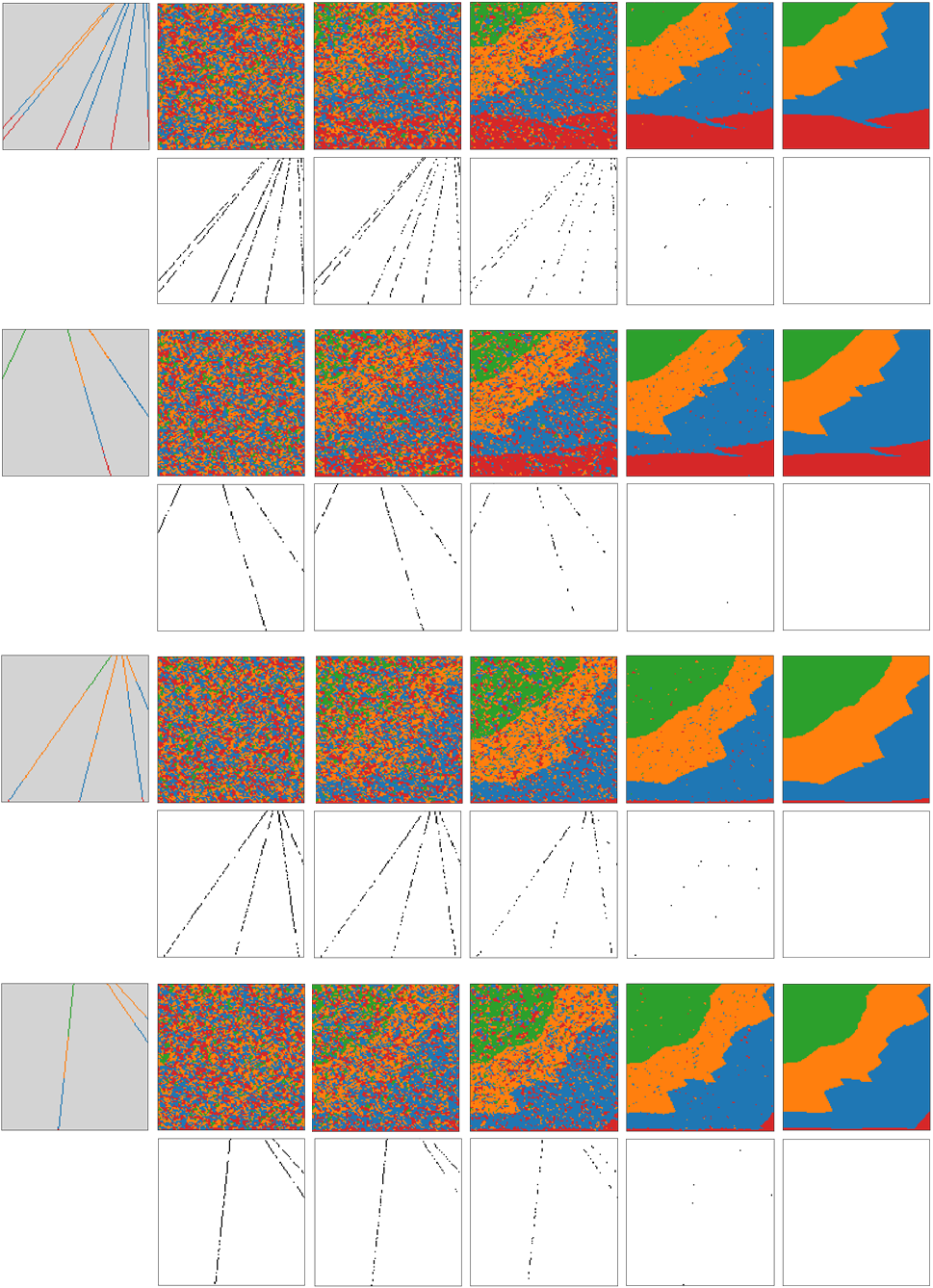}
  \caption{(continuation of Fig.~\ref{fig:transitional_sampling_ldm}) Additional examples of the transitions in the conditional denoising process.
  }
  \label{fig:transitional_sampling_ldm-2}
\end{figure}

\end{document}